\documentclass[sigconf,natbib=true]{acmart}

\usepackage{multirow,makecell}
\usepackage{soul}
\usepackage{graphicx, caption, subcaption, amsmath}
\usepackage{enumitem}

\usepackage{array}
\usepackage{booktabs}
\usepackage{bbding}
\usepackage{threeparttable}
\usepackage{algpseudocode}
\usepackage{bm}

\newcolumntype{L}[1]{>{\raggedright\arraybackslash}p{#1}}
\newcolumntype{C}[1]{>{\centering\arraybackslash}p{#1}}
\newcolumntype{R}[1]{>{\raggedleft\arraybackslash}p{#1}}

\setlist[itemize]{leftmargin=*}

\definecolor{blue}{RGB}{113,168,207}
\definecolor{orange}{RGB}{255, 127, 14}

\AtBeginDocument{%
  \providecommand\BibTeX{{%
    \normalfont B\kern-0.5em{\scshape i\kern-0.25em b}\kern-0.8em\TeX}}}

\copyrightyear{2024}
\acmYear{2024}
\setcopyright{rightsretained}
\acmConference[SIGIR '24] {Proceedings of the 47th International ACM SIGIR Conference on Research and Development in Information Retrieval}{July 14--18, 2024}{Washington, DC, USA.}
\acmBooktitle{Proceedings of the 47th International ACM SIGIR Conference on Research and Development in Information Retrieval (SIGIR '24), July 14--18, 2024, Washington, DC, USA}
\acmISBN{979-8-4007-0431-4/24/07}
\acmDOI{10.1145/3626772.3657777}

\usepackage{etoolbox}
\makeatletter
\patchcmd{\maketitle}{\@copyrightpermission}{
  \begin{minipage}{0.3\columnwidth}
     \href{https://creativecommons.org/licenses/by/4.0/}{\includegraphics[width=0.90\textwidth]{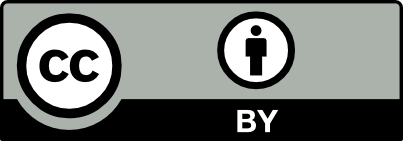}}
  \end{minipage}\hfill
  \begin{minipage}{0.7\columnwidth}
     \href{https://creativecommons.org/licenses/by/4.0/}{This work is licensed under a Creative Commons Attribution International 4.0 License.}
  \end{minipage}
  
  \vspace{5pt}
}{}{}

\makeatother

\begin{document}

\title{Deep Pattern Network for Click-Through Rate Prediction}

\author{Hengyu Zhang}
\email{zhang-hy21@mails.tsinghua.edu.cn}
\affiliation{
  \institution{Tsinghua Shenzhen International Graduate School, Tsinghua University}
  \city{Shenzhen}
  \country{China}}

\author{Junwei Pan}
\email{jonaspan@tencent.com}
\affiliation{
  \institution{Tencent}
  \city{Shenzhen}
  \country{China}}

\author{Dapeng Liu}
\email{rocliu@tencent.com}
\affiliation{
  \institution{Tencent}
  \city{Shenzhen}
  \country{China}}

\author{Jie Jiang}
\email{zeus@tencent.com}
\affiliation{
  \institution{Tencent}
  \city{Shenzhen}
  \country{China}}

\author{Xiu Li}
\email{li.xiu@sz.tsinghua.edu.cn}
\authornote{Corresponding author.}
\affiliation{
  \institution{Tsinghua Shenzhen International Graduate School, Tsinghua University}
  \city{Shenzhen}
  \country{China}}

\begin{abstract}

Click-through rate (CTR) prediction tasks play a pivotal role in real-world applications, particularly in recommendation systems and online advertising. A significant research branch in this domain focuses on user behavior modeling. Current research predominantly centers on modeling co-occurrence relationships between the target item and items previously interacted with by users in their historical data. However, this focus neglects the intricate modeling of user behavior patterns.
In reality, the abundance of user interaction records encompasses diverse behavior patterns, indicative of a spectrum of habitual paradigms. These patterns harbor substantial potential to significantly enhance CTR prediction performance. To harness the informational potential within user behavior patterns, we extend Target Attention (TA) to Target Pattern Attention (TPA) to model pattern-level dependencies.
Furthermore, three critical challenges demand attention: the inclusion of unrelated items within behavior patterns, data sparsity in behavior patterns, and computational complexity arising from numerous patterns. To address these challenges, we introduce the Deep Pattern Network (DPN), designed to comprehensively leverage information from user behavior patterns. DPN efficiently retrieves target-related user behavior patterns using a target-aware attention mechanism. Additionally, it contributes to refining user behavior patterns through a pre-training paradigm based on self-supervised learning while promoting dependency learning within sparse patterns.
Our comprehensive experiments, conducted across three public datasets, substantiate the superior performance and broad compatibility of DPN.

\end{abstract}


\maketitle

\section{Introduction}
\label{intro}

Click-Through Rate (CTR) prediction holds paramount significance in industrial scenarios such as online advertising and recommender systems, where the goal is to predict the probability of the specific user clicking on a target item.
Effective CTR prediction not only serves to maximize revenue for advertisers but also enhances the user experience by delivering more pertinent content.
The pursuit of better prediction models has led to the widespread adoption of deep learning techniques in the field of click-through rate prediction.

Deep-learning-based CTR prediction methods have taken the forefront in current research and achieved remarkable success~\cite{dl4ctr}, which aim to capture intricate feature interactions using neural networks.
User behavior modeling, which seeks to extract latent interest from the extensive user history behavior records via various techniques including pooling, RNN-based~\cite{gru4rec, dien}, CNN-based~\cite{caser}, Graph-based~\cite{beyondclick,NICTR,hpmr}, and attention~\cite{sasrec, bert4rec, bst, dualrec} model, represent one of the most crucial research branches~\cite{ubm, din, dien, tin}.

\begin{figure}[!ht]
\centering
    \includegraphics[width=0.90\linewidth]{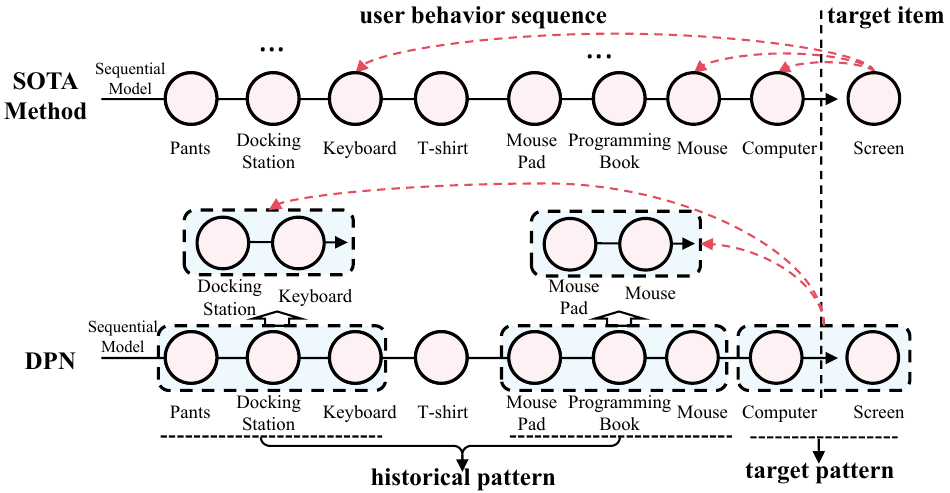} 
    \caption{Illustration of the dependency modeling that the existing SOTA methods and DPN focus on. The red dotted line indicates Target Attention or Target Pattern Attention.}
    \label{fig:intro2}
\end{figure}

DIN~\cite{din} stands out as a notable milestone in user behavior modeling. 
User interest is diverse and varies for different target items.
DIN handles this issue by introducing a target-attention mechanism that enables the extraction of specific interests for a given target item.
Due to its superiority, DIN-based methods have become the mainstream branch in CTR prediction based on user behavior modeling. 
DIEN~\cite{dien} further improves the model performance by incorporating GRU~\cite{gru} networks for dynamic evolutionary modeling.
Recent research endeavors have extended this progress in terms of modeling long sequences of user behavior~\cite{mimn, sim, eta} and co-occurrence behavior representation modeling~\cite{can}.
In summary, the state-of-the-art (SOTA) methods typically apply a sequential model to handle the user behavior sequence and use target attention to aggregate the item representations, as shown in Figure \ref{fig:intro2}.

\begin{figure}[!h]
\centering
\begin{subfigure}{0.4955\linewidth}
    \centering
    \includegraphics[width=1.05\linewidth]{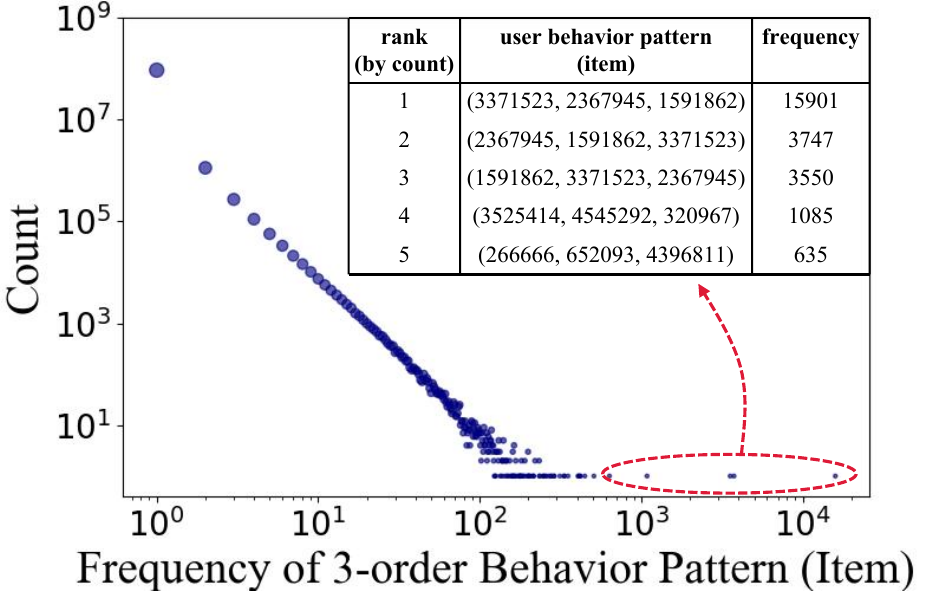} 
\end{subfigure}
\begin{subfigure}{0.4955\linewidth}
    \centering
    \includegraphics[width=1.05\linewidth]{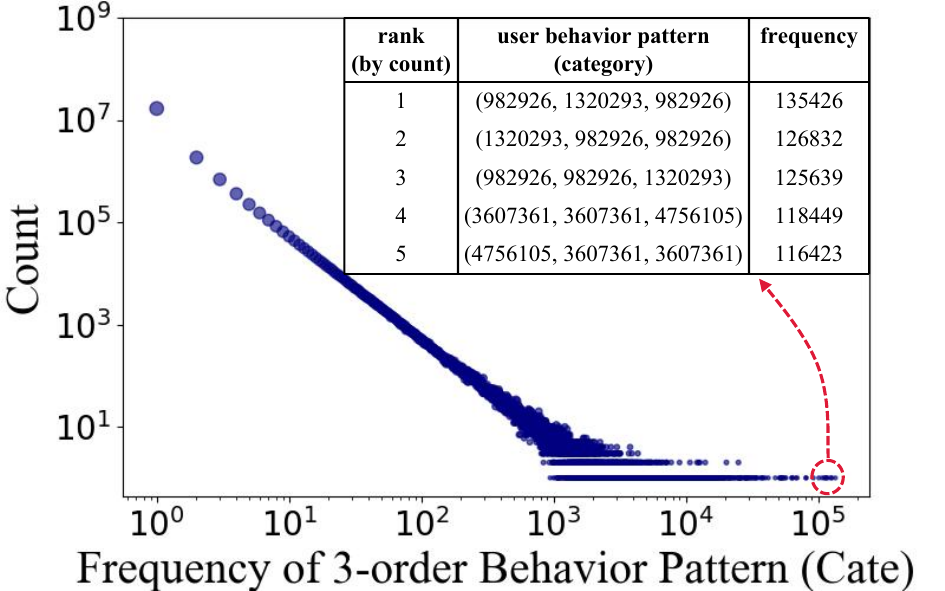} 
\end{subfigure}
    \caption{Statistical results for 3-order user behavior patterns in the Taobao dataset (log-scale plot). Each scatter point with coordinates $(x,y)$ represents that the number of 3-order user behavior patterns with occurrence frequency $x$ in the Taobao dataset is $y$. The figure shows the Top-5 high-frequency user behavior patterns for items and categories, respectively.}
\label{fig:intro}
\end{figure}

According to psychological studies~\cite{psy1,psy2}, the entire user personality is linked to a variety of behavior patterns.
Specifically, a \textit{behavior pattern} can be defined as a \textit{subsequence of two or more actions that occur in a prescribed arrangement}, such as purchasing a \textit{mouse pad} and then buying a \textit{mouse} as shown in Figure \ref{fig:intro2}.
The user behavior sequence is composed of diverse behavior patterns driven by varying user interest.
However, the existing SOTA methods uniformly model user behavior sequences containing diverse interests via a unified sequential model, overlooking the intricate modeling of the varied behavior patterns within. 
These methods solely focus on item-level dependency, neglecting the critical aspect of dependency modeling among patterns.
Furthermore, many fixed behavior patterns frequently appear in recommendation scenarios as shown in Figure \ref{fig:intro}, demonstrating they embed meaningful dependency information.
Hence, exploring how to effectively leverage diverse behavior patterns in CTR prediction tasks is an intriguing research question.

To fully leverage the information from behavior patterns, we extend Target Attention (TA) into \textbf{T}arget \textbf{P}attern \textbf{A}ttention (\textbf{TPA}) to perform \textit{pattern-level} interest aggregation. TPA models the dependency between the target behavior pattern and the historical patterns, where the target pattern consists of the user's recent behaviors and the interaction to be predicted with the target item. However, due to the multitude and complexity of behavior patterns contained within user behavior sequences, there are several challenges making it a non-trivial problem:

\begin{itemize}
    
    \item \textbf{\textit{C1:} Unrelated items mixed into behavior patterns.}
    In some cases, behavior patterns may not be continuous segments, which means some irrelevant items misclicked or driven by other interest may mix in.
    For example in Figure \ref{fig:intro2}, the user interacts with the mouse pad, programming book, and mouse successively, but the programming book may not have a strong relationship with the pattern (mouse pad, mouse).
    So effective refinement of behavior patterns is necessary, which guarantees patterns to be more meaningful, i.e., with stronger intra-dependencies within the items in behavior patterns, and reduces the risks of introducing noise.
    
    \item \textbf{\textit{C2:} Data Sparsity of Behavior Patterns.}
    Behavior patterns are subsequences of multiple items interacted with by the user, implying high-order item dependencies. 
    Some behavior patterns exhibit high sparsity, leading to difficulty in dependency learning within them.

    \item \textbf{\textit{C3:} Computational
complexity arising from numerous patterns.}
    User historical behavior records contain a plethora of complex user behavior patterns driven by different interest.
    Modeling all the behavior patterns incurs unacceptable computation costs, and the behavior patterns irrelevant to the target item may introduce noises.
    Consequently, efficiently retrieving target-related behavior patterns from the abundance of user historical behaviors is a crucial issue to address.
\end{itemize}

To address the challenges above, we introduce a novel \textbf{D}eep \textbf{P}attern \textbf{N}etwork (\textbf{DPN}) for the Click-Through Rate prediction task.
In contrast to existing user behavior modeling methods, DPN distinguishes itself by not only efficiently discovering effective behavior patterns but also introducing pattern-level dependency modeling, as illustrated in Figure \ref{fig:intro2}.

Firstly, for the efficiency of behavior pattern modeling (\textbf{\textit{C3}}), we introduce a Target-aware Pattern Retrieval Module (TPRM) to identify the Top-K target-related user behavior patterns.
Secondly, we ingeniously integrate the two major desideratas of effectively refining behavior patterns (\textbf{\textit{C1}}) and addressing pattern sparsity (\textbf{\textit{C2}}). 
To achieve this, we design an innovative self-supervised pattern refinement module, thereby avoiding excessive time and storage overhead associated with complex structural designs.
SPRM is pretrained by a self-supervised denoising task, specifically removing noise introduced by random augmentation of behavior patterns, enhancing the process of pattern refinement and optimizing pattern representation learning.
Thirdly, DPN extends the concept of target attention to pattern level, proposing Target Pattern Attention to model the dependencies between patterns.

From a model expansion standpoint, our DPN can be regarded as \textbf{extend} the width of Query, Key, and Value within the attention mechanism, providing a new perspective that deviates significantly from previous efforts concentrated on making attention module in recommendation longer~\cite{sim, eta, sdim} or deeper~\cite{HSTU}.
This idea shares some similarities with works in CV~\cite{ViT,RepLKNet} and NLP~\cite{zen,n-gram}. 
For instance, the concept of widening query, key, and value from a single item to the pattern (consecutive items) resembles extending convolution windows from $1 \times 1$ to larger ones, enhancing the model's perception of the local context. 
To some extent, our work can also be aligned with Vision Transformers (ViT), where items can be regarded as pixels and patterns as patches.
In the realm of NLP, some works like ZEN~\cite{zen} also explore the expansion of models from uni-gram to N-gram to enhance transformer representations.

Our contributions can be summarized as follows:

\begin{itemize}
    \item We highlight the significance of behavior patterns and summarize the challenges of efficiently exploiting them.
    \item To address these challenges, we propose DPN, which includes target-aware retrieval, self-supervised refinement, and pattern-level interest aggregation to fully leverage behavior patterns.
    \item We conduct comprehensive experiments on three public real-world datasets to demonstrate the superior effectiveness and broad compatibility of our proposed DPN. Furthermore, further in-depth analysis of how user behavior patterns are utilized is conducted.
\end{itemize}

\begin{figure*}[!ht]
    \centering
    \setlength{\abovecaptionskip}{1mm}
    \setlength{\belowcaptionskip}{0mm}
    \includegraphics[width=0.86\linewidth]{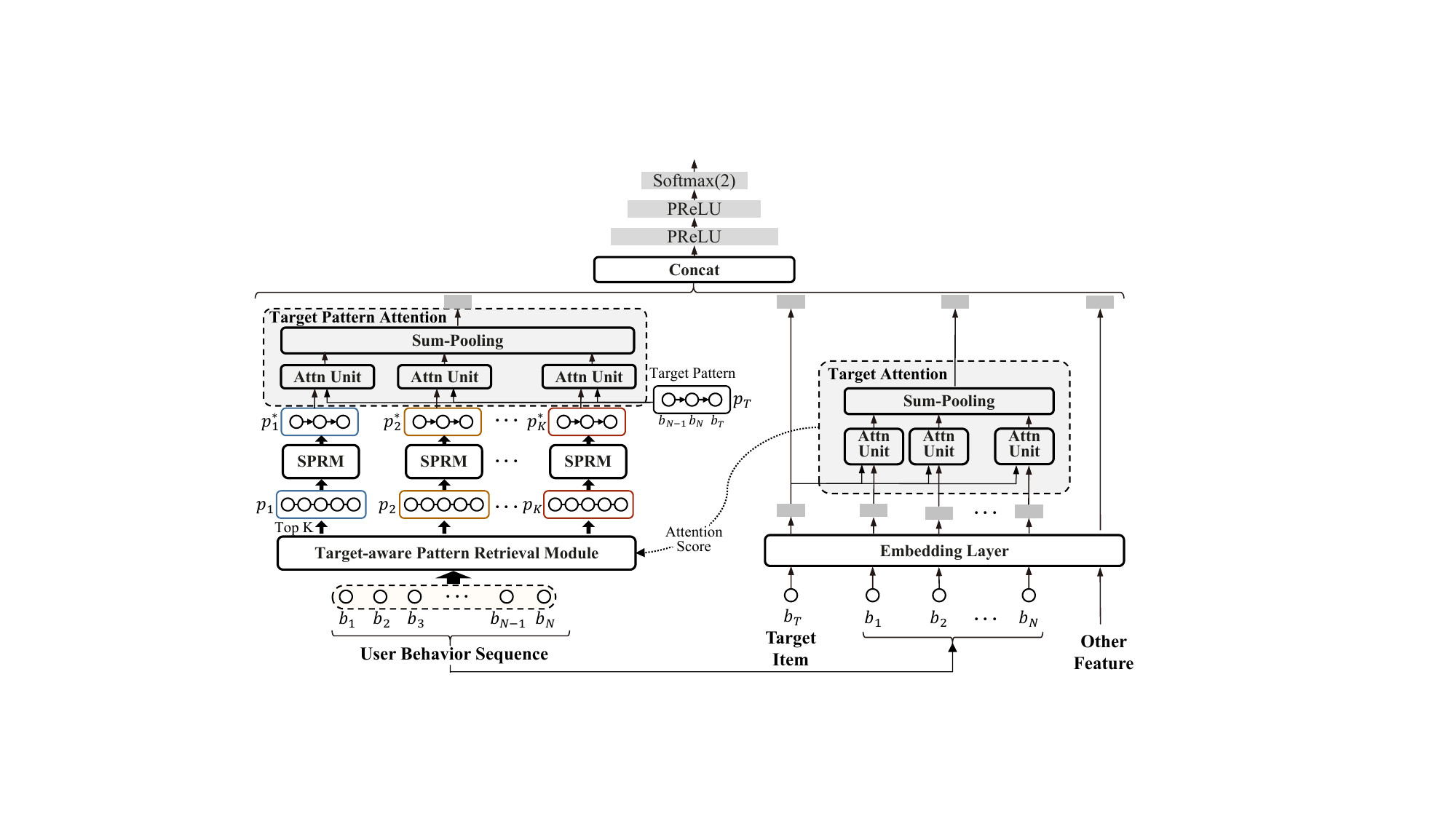}
    \caption{Overall architecture of the proposed DPN. Target-ware Pattern Retrieval Model (TPRM) searches the Top-K target-related pattern from user behavior sequence according to target attention scores. SPRM adopts a pre-trained refinement network to perform fine-grained denoising of the behavior patterns, whose detailed illustration is shown in Figure \ref{fig:SPRM}. TPA models the dependency between the target behavior pattern and refined historical behavior patterns.}

\label{fig:model}
\end{figure*}

\section{Related Works}

In early research work, click-through rate prediction models mainly focused on improving CTR prediction performance by means of feature engineering~\cite{poly2,FM,mcmahan2013ad}.
In recent years, deep learning techniques have developed rapidly and achieved impressive results in areas such as computer vision~\cite{vgg,resnet} and natural language processing~\cite{transformer,seq2seq}.
Therefore, researchers are currently working on capturing feature interactions automatically through deep neural networks.
Approaches like DeepFM~\cite{deepfm} and its derivatives xDeepFM~\cite{xdeepfm} employ neural networks inspired by factorization machines to autonomously learn feature interactions, eliminating the need for manual feature engineering.
PNN~\cite{pnn} introduces a product layer capable of capturing feature interactions by utilizing either inner or outer products as factorization functions. 
Lastly, ONN~\cite{onn} devises operation-aware embedding layers to enhance feature interaction through enriched embeddings.

A significant research branch in CTR prediction focuses on user behavior modeling, giving birth to a series of representative works~\cite{ubm, din, dien, can}.
The CTR prediction model based on user behavior modeling is dedicated to capturing the implicit interest representations of users from the rich history of their behavior records.
DIN~\cite{din} is a pioneer work in this field of research.
It proposes a target attention mechanism, which treats target items as queries and user history as keys and values, to aggregate the user's specific interests for a given target item from the user's history of interactions.
DIEN~\cite{dien}, an improved version of DIN, captures dynamic interest representations of users by modeling the evolution of their interests through GRU~\cite{gru}.
AutoAttention~\cite{autoattention} automated field selection in the target attention mechanism.
MIMN~\cite{mimn} employs a memory network~\cite{ntm} to consolidate historical user behavior data, effectively tackling the challenge of modeling users' long-term interests.
Subsequent research, exemplified by SIM~\cite{sim} and ETA~\cite{eta}, concentrates on the efficient retrieval-based modeling of long-term interests.
CAN~\cite{can} disentangles the process of modeling feature interactions from the initial feature modeling, by directing features into a compact multi-layer perceptron (mini-MLP)  generated by other features.
TIN~\cite{tin} incorporates target-aware temporal encoding into the target attention mechanism to capture both semantic and temporal correlation.

\section{Method}
\label{sec:Model}

The overall architecture of DPN is depicted in Figure \ref{fig:model}.
The inputs to DPN encompass the target item to be predicted in the CTR prediction task, the user's historical behavior sequence, and other relevant features.
The goal of DPN is to predict the probability of a user interacting with the target item.

\subsection{Base Model}
\label{sec:base}

\subsubsection{Embedding Layer}

For a given set of users $\mathcal{U}$, the historical behavior records of user $u \in \mathcal{U}$ can be formulated as a chronological sequence $S^u = \{b^u_i, b^u_2, \cdots, b^u_{N_u}\}$, where $b^u_k$ denotes the $k$-th behavior record of user $u$ and $N_u$ denotes the length of user interaction sequence. 
Notably, the superscript $u$ will be omitted for simplicity in the following presentation. 
To enable parallel computation in tensor form, the behavior sequence should be truncated or padded to a fixed length $N$, resulting in a fix-length behavior sequence $S=\{b_1,b_2,\cdots,b_{N}\}$.

Typically, the behavior record $b_k$ consists of several features, such as item ID $i_k$ and category ID $c_k$ for our experiments. 
So for each interaction $b_k$, we can map it to a dense embedding $\mathbf{e}_k$ by concatenating the embeddings of corresponding features obtained via LookUp tables, which can be formulated as follows:
\begin{equation}
    \mathbf{e}_k = \left[\operatorname{LookUP}(i_k, \mathbf{E}^{\mathcal{I}}); \operatorname{LookUP}(c_k, \mathbf{E}^{\mathcal{C}})\right],
\end{equation}
where $[\cdot;\cdot]$ means concatenateing the embedding vectors, and $\operatorname{LookUp}(a, \mathbf{E})$ denotes the operation of retrieving the $a$-th embedding vector in embedding matrix $\mathbf{E}$. $\mathbf{E}^{\mathcal{I}} \in \mathbb{R}^{|\mathcal{I}|\times d}, \mathbf{E}^{\mathcal{C}} \in \mathbb{R}^{|\mathcal{C}|\times d}$ are embedding matrixes of items and categories, respectively, where $d$ denotes the embedding dimensionality and $|\mathcal{I}|, |\mathcal{C}|$ means the volume of item space and category space.

Thus,  we can get the sequence of user behavior embeddings:
\begin{equation}
    \mathbf{E}_{S}=\{\mathbf{e}_1, \mathbf{e}_2, \cdots, \mathbf{e}_N\}.
\end{equation}
\subsubsection{User Interest Extraction Module}

For a fair comparison, we adopt the typical user-behavior-based CTR prediction model DIN 
\cite{din} as the user interest extraction backbone, following~\cite{sim, eta}.
DIN proposes the target attention mechanism to capture the specific user interest representation $\mathbf{v}_T$ for the target item $i_T$ of the target category $c_T$.

The target attention mechanism can be formulated as:
\begin{equation}
    \mathbf{v}_T = \sum\limits_{k=1}^{N}a(\mathbf{e}_T, \mathbf{e}_k) \cdot \mathbf{e}_k,
\end{equation}
where $a(\cdot,\cdot)$ means the attention function, $\mathbf{e}_T$ denotes the embedding of the target item $\mathbf{i}_T$ and the target category $\mathbf{c}_T$, which is to be predicted whether to be clicked by the user. The attention function used in DIN~\cite{din} is $a(\mathbf{q}, \mathbf{k}) = \operatorname{softmax}\left(\operatorname{MLP}(\left[\mathbf{q}; \mathbf{k}; \mathbf{q}-\mathbf{k}; \mathbf{q}\circ\mathbf{k}\right])\right)$, where $\circ$ denotes the Hadamard product, i.e., element-wise multiplication. 

\subsection{Target-aware Pattern Retrieval Module (TPRM)}

Target-aware Pattern Retrieval Module aims to search for the user behavior pattern relative to the target item and category.

We retrieve the positions with Top-K attention scores in the User Interest Extraction Module to obtain similar behavior records with the target item $i_T$ and category $c_T$, which can be presented as follows:
\begin{equation}
    idx_1, idx_2,\cdots,idx_K=\mathop{\mathrm{argtopk}}\limits_K\  a(\mathbf{e}_T, \mathbf{e}_1),\cdots, a(\mathbf{e}_T, \mathbf{e}_N),
\end{equation}
where $\mathop{\mathrm{argtopk}}\limits_K$ retrieves the corresponding indexes of the Top-K largest attention scores.
The retrieval method is not the core of our work, it can be replaced by other approaches such as locality-sensitive hashing for ultra-long sequence modeling.

The $k$-th user behavior pattern $p_k$ of length $l$ consists of the $idx_k$-th interaction and the $l-1$ interactions preceding it, which can be denoted as:
\begin{equation}
    p_k=\{b_{idx_k-l+1},\cdots,b_{idx_k-1},b_{idx_k}\}, \quad k=1,2,\cdots,K.
\end{equation}

As a result, we obtain $K$ user behavior patterns. Since this is a target-aware retrieval process, these retrieved behavior patterns reflect the behavior paradigms that users may engage in before clicking on the target item and category.

\subsection{Self-supervised Pattern Refinement Module (SPRM)}

\begin{figure}[!htbp]
    \centering
    \includegraphics[width=\linewidth]{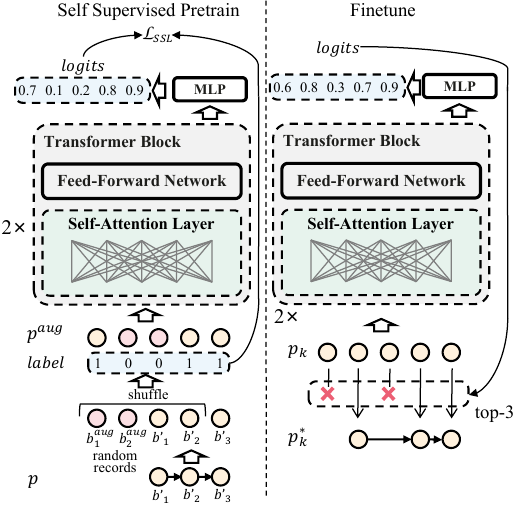}
    \caption{Illustration of Self-supervised Pattern Refinement Module (SPRM). On the left, random data augmentation and denoising objectives are applied to pretrain refinement network based on self-supervised learning. On the right, the refinement network is utilized to generate a logit vector for pattern refinement in CTR prediction tasks.}
\label{fig:SPRM}
\end{figure}

User behavior patterns are not always continuous segments, which means some irrelevant items misclicked or driven by other interest may mix in.
For example, if the user interacts with the iPhone, T-shirt, and Airpods successively, only the transition from iPhone to Airpods is a meaningful pattern.
Therefore, we propose a self-supervised pattern refinement network to extract true user behavior patterns from the retrieved raw user behavior patterns, as presented in Figure \ref{fig:SPRM}.
Firstly, the refinement network is pretrained by a self-supervised denoising task.
Then DPN applies the pretrained refinement network to refine the retrieved pattern to be more meaningful, which also ensures that subsequent pattern-level dependency modeling is more meaningful.

\textbf{Data Augmentation.}

We take sequential interaction records in the dataset, i.e., one user behavior pattern, as the raw input $p=\{b'_1,b'_2,\cdots,b'_s\}$. We then perform data augmentation by randomly mixing in interaction records with some sampled interaction records into the original behavior pattern, resulting in an augmented pattern $p^{aug}=\{b''_1, b''_2, \cdots, b''_l\}$. Notably, the last record of the augmented behavior pattern sequence and the original behavior pattern sequence should be consistent, which indicates to the refinement network what the decision goal of the behavior pattern is.

\textbf{Refinement Network.}

We use a two-layer Transformer network as an encoder to extract behavior pattern representations from the augmented sequence of user behavior patterns. For a specific augmented behavior pattern $p^{aug}=\{b''_1, b''_2, \cdots, b''_l\}$, the encoder can be formalized as follows:
\begin{equation}
    \mathbf{X}^{(0)} = \left[\mathbf{e}''_1+\mathbf{e}^{pos}_1, \mathbf{e}''_2+\mathbf{e}^{pos}_2, \cdots, \mathbf{e}''_l+\mathbf{e}^{pos}_l\right],
\end{equation}
\begin{equation}
    \begin{aligned}
    \mathbf{H}^{(m)}&=\operatorname{LayerNorm}\left(\mathbf{X}^{(m-1)}+\operatorname{SA}\left(\mathbf{X}^{(m-1)}\right)\right),\\
    \mathbf{X}^{(m)}&=\operatorname{LayerNorm}\left(\mathbf{H}^{(m)}+\operatorname{FFN}\left(\mathbf{H}^{(m)}\right)\right),
    \end{aligned}
\end{equation}
where $\mathbf{e}''_j$ and $\mathbf{e}^{pos}_j$ are the embedding vectors of $j$-th record $b''_j$ in augmented behavior pattern $p^{aug}$ and the $j$-th positional embedding, respectively. 
$\operatorname{SA}$ and $\operatorname{FFN}$ denote the Self-Attention Layer and Feed-Forward Network, $\operatorname{LayerNorm}$ means layer normalization.
$\mathbf{X}^{(m)}, \mathbf{H}^{(m)}$ means the hidden representation and the intermediate representation at layer $m$, and $\mathbf{X}^{(0)}$ is the input of transformer.

Thus, we can get the representation vector $\mathbf{X}^{(M)}$ of the augmented behavior pattern $p^{aug}$, where the depth $M$ of the Transformer network is set to 2.
Then, we encode the representation into a logits vector $\mathbf{o}$ of length $l$ by MLP to predict which interaction records in the augmented behavior pattern $p^{aug}$ correspond to the raw behavior pattern $p$, which can be formulated as:
\begin{equation}
    \mathbf{o} = \operatorname{MLP}\left(\mathbf{X}^{(M))}\right) \in \mathbb{R}^{l},
\end{equation}
where the positions corresponding to the Top-$s$ largest values in logits vector $\mathbf{o}$ indicate the predicted interaction records belonging in raw pattern $p$.
In the CTR prediction task, SPRM selects Top-$s$ interaction records according to the logits vector $\mathbf{o}$ as the refined behavior pattern, as shown in the right part of Figure \ref{fig:SPRM}.

\textbf{Loss Function.}

Cross-entropy loss is used for pre-training based on self-supervised learning to guide the model in the refinement of behavior patterns.
The loss function can be represented as:
\begin{equation}
    \mathcal{L}_{SSL}=-\sum\limits_{j=1}^{l} \mathbb{I}_{p}(b''_j)*\log (o_j), b''_j \in p^{aug},
\end{equation}
where $\mathbb{I}_{p}(b''_j)$ denotes the indicator function to point out whether $b''_j$ belongs to behavior pattern $p$. 
If so, the function output is 1, and vice versa 0.

\subsection{Target Pattern Attention (TPA)}

User behavior patterns reflect the implicit psychological decision paradigm of users. Modeling the dependency between the historical behavior patterns and the current target behavior pattern can effectively help the model determine whether the clicks on the target items and categories are in line with the user's habitual paradigm, thus improving the model's recommendation performance.

DPN extends the well-known target attention to Target Pattern Attention to achieve inter-pattern dependency modeling.
After obtaining the refined target-related user behavior patterns $p^*_k=\{b^*_1, b^*_2, \cdots, b^*_s\}$, DPN applies the attention units to capture the dependency between the refined  pattern $p_k$ and the target behavior pattern $p_T=\{b_{N-s+2}, \cdots, b_{N},b_T\}$ resulting in pattern-level interest representation $\mathbf{v}_p$ that are used in the final CTR prediction task:
\begin{equation}
    \mathbf{v}_p = \sum\limits_{k=1}^K a(\mathbf{E}_{p_T}, \mathbf{E}_{p^*_k})\cdot\mathbf{E}_{p^*_k},
\end{equation}
where $\mathbf{E}_{p^*_k}$ and $\mathbf{E}_{p_T}$ denote the representations of $p^*_k$ and $p_T$ generated by Transformer encoders, respectively.

\subsection{Training Objective}

The task of click-through rate prediction can be structured as a binary classification problem, specifically, determining whether the target instance will be clicked or not. Consequently, the final output of the DPN model can be expressed as:
\begin{equation}
    (\mathop{\mathop{\hat y}\limits_{\uparrow}}_{\text{click}}, \mathop{\mathop{1-\hat y}\limits_{\uparrow}}_{\text{unclick}}) = \operatorname{softmax}(\operatorname{MLP}(\left[\mathbf{e}_T;\sum\limits_{j=1}^N\mathbf{e}_{j}; \mathbf{v}_T; \mathbf{v}_p\circ\mathbf{E}_{p_T}\right]).
\end{equation}
Here, $\hat y$ and $1-\hat y$ represent the predicted logit values for click and unclick, respectively. $\operatorname{MLP}$ denotes a compact neural network with three layers and a 2-dimensional output.

The training objective of the CTR prediction task can be defined as follows:
\begin{equation}
    \min_{\Theta} \ \mathcal{L}_{CTR} = -\frac{1}{B}\sum\limits_{\mathcal{B}}\left(y\cdot\log(\hat y) + (1-y)\cdot\log(1-\hat y)\right).
\end{equation}

In this equation, $y$ is the label indicating whether the user will interact with the target item.
$\Theta$ denotes the set of trainable parameters in DPN, and $\mathcal{B}$ represents a batch of instances with a batch size of $B$.

\subsection{Model Discussion}

\subsubsection{Time Complexity Analysis}

DPN mainly contains the Base Model and three core components.
The time complexity of the Base Model, which defaults to DIN, is $\mathcal{O}(BNd)$.
TPRM is proposed to retrieve Top-K target-related behavior patterns from user behavior sequences, so its time complexity is $\mathcal{O}(BN\log K)$ if Top-K retrieval is performed using heap structure.
A two-layer Transformer is used in SPRM to encode the raw user behavior patterns.
The time complexity of self-attention layer and feed-forward network are $\mathcal{O}(BKl^2d)$ and $\mathcal{O}(BKld^2)$, respectively, so the complexity amount of SPRM is $\mathcal{O}(BKld(l+d))$.
TPA involves only attention units between historical behavior patterns and target behavior patterns, which has a time complexity of $\mathcal{O}(BKd)$.

\section{Experiments}

\subsection{Experiment Settings}
\begin{table}[!ht]
\caption{Statistics of evaluation datasets.}
\centering
\begin{tabular}{C{1.5cm}|C{1.5cm}C{1.5cm}C{1.5cm}}
\toprule
Dataset & \#User & \#Item & \#Interactions  \\ \hline
Tmall & 424,170&  1,090,390 &  54,925,330\\ \hline
Taobao & 987,994 & 4,162,024 & 100,150,807\\\hline
Alipay & 963,923& 2,353,207& 44,528,127\\\bottomrule
\end{tabular}
\label{tab:dataset}
\end{table}

\subsubsection{Dataset Descriptions.}

To comprehensively assess the performance of our proposed DPN, we conducted experiments on three public recommendation datasets. These datasets are as follows:

\begin{itemize}
    \item \textbf{Tmall Dataset\footnote{https://tianchi.aliyun.com/dataset/dataDetail?dataId=42}.} This dataset comprises user behavior records on Tmall.com, spanning from May 2015 to November 2015. To construct the user behavior sequences for our experiments, we arrange the interaction records from the Tmall dataset in chronological order based on their action time.

    \item \textbf{Taobao Dataset\footnote{https://tianchi.aliyun.com/dataset/dataDetail?dataId=649}.} This dataset, provided by Alibaba Group, captures diverse user shopping activities on Taobao.com, including clicks, adding items to the cart, and making purchases. In our experiments, we treat all these user behaviors as interactions.

    \item \textbf{Alipay Dataset\footnote{https://tianchi.aliyun.com/dataset/dataDetail?dataId=53}.}
    Alipay dataset encompasses a comprehensive record of user activities both online and on-site during the period spanning from July 1st, 2015, to November 30th, 2015. This extensive dataset draws its data from Tmall.com, Taobao.com, and the Alipay app, providing valuable insights into user behaviors and interactions with merchants.
    
\end{itemize}

\subsubsection{Evaluation Metrics.}

In our research trials, we assess the effectiveness of the models by employing the Area Under the Curve (AUC) metric, which is the most frequently used metric in Click-Through Rate (CTR) prediction tasks. AUC gauges the model's ability to prioritize positive samples over randomly selected negative samples. A greater AUC value signifies improved performance in predicting CTR.

For a prediction model with random parameters, the model's AUC value is approximately 0.5. Therefore, the worst-case scenario for AUC is not 0, as with other metrics, but rather 0.5. This is why its relative gain is somewhat different. The definition of relative improvement of AUC is as follows~\cite{din, relaimpr}:
\begin{equation}
    RelaImpr. = \left(\frac{\text{AUC}-0.5}{\text{AUC}_{base}-0.5}-1\right)\times 100\%,
    \label{relaimprov}
\end{equation}

\subsubsection{Comparision Baselines.}

In order to thoroughly assess the effectiveness of our DPN proposal, we conduct a comprehensive comparison with eleven state-of-the-art CTR prediction baseline models, including \textbf{DNN}~\cite{dnn}, \textbf{NCF}~\cite{ncf}, \textbf{Wide\&Deep}~\cite{widedeep} , \textbf{PNN}~\cite{pnn},  \textbf{RUM}~\cite{rum},  \textbf{ONN}~\cite{onn}, \textbf{DIN}~\cite{din}, \textbf{GRU4Rec}~\cite{gru4rec},  \textbf{DIEN}~\cite{dien}, \textbf{MIMN}~\cite{mimn}, \textbf{SIM}~\cite{sim}, \textbf{ETA}~\cite{eta}, \textbf{CAN}~\cite{can}.

\subsubsection{Implement Details.}

All the baseline techniques and our proposed DPN are implemented using the TensorFlow framework~\cite{TensorFlow}. 
The source code for DPN will be available soon.
To ensure a fair comparison, we establish the following parameter settings: a maximum sequence length of $N=100$ and an embedding dimensionality of $d=16$ for items and categories. The number of items retrieved by the retrieval-enhanced methods (SIM, ETA) is aligned with the number of items contained in utilized patterns in DPN, i.e., $K\times s$. We employ the Adam optimizer~\cite{adam} with a learning rate of 0.001 for training all methods. The batch size is set to 256 for both the training dataset and the testing dataset.

During the pretraining process of SPRM, we aligned the lengths of the behavior patterns before and after augmentation, denoted as $l$ and $s$ respectively, with the settings of the CTR prediction task. We employed the Adam optimizer with a learning rate of 0.001 for the pretraining phase. The batch size was set to 8196, and the entire pretraining process comprised 10 epochs.

Regarding the hyperparameters for DPN, we have configured the number of retrieved behavior patterns in TPRM, $K$, to be 5, and the length of the retrieved behavior patterns, $l$, to be 5. The length of refined behavior patterns in SPRM, $s$, is set to 3.

\subsection{Overall Performances}

\begin{table}[!ht]
\caption{
Overall performance of various methods on three publicly available recommendation datasets, the most effective method is highlighted in bold. An asterisk (*) is used to indicate the statistical significance (with p < 0.05) when comparing DPN to the top-performing baseline results.}
\begin{tabular}{L{2.6cm}|C{1.3cm}C{1.3cm}C{1.3cm}}
\toprule
\multirow{2}{*}{Method} & \multicolumn{3}{c}{AUC} \\\cline{2-4}
                        & Tmall    & Taobao   &   Alipay\\\hline
DNN(Sum-Pooling)                     & 0.9268   & 0.8731   &   0.9037\\
NCF                     & 0.9256   & 0.8762   &   0.9046\\
Wide\&Deep              & 0.9362   & 0.8751   &   0.9040\\
PNN                     & 0.9408   & 0.8839   &   0.9058\\
RUM                     & 0.9386   & 0.9046   &   0.9232\\
GRU4REC                 & 0.9392   & 0.9056   &   0.9191\\
DIN                     & 0.9307   & 0.8931   &   0.9185\\
DIEN                    & 0.9459   & 0.9058   &   0.9251\\
ONN                     & 0.9459   & 0.9062   &   0.9122\\
MIMN                    & 0.9458   & 0.9162   &   0.9280\\
SIM                    & 0.9494	   & 0.9281	  &   0.9205\\
ETA                    & 0.9466	   & 0.9259	  &   0.9202\\
CAN                     & 0.9504   & 0.9327   &   0.9311\\\hline
DPN (Ours)             &\textbf{0.9573$^*$}&          \textbf{0.9431$^*$}&          \textbf{0.9438$^*$}  \\\bottomrule
\end{tabular}
\label{tab:overall}
\end{table}

The comprehensive performance evaluation of our novel DPN is presented in Table \ref{tab:overall}. 
Based on the findings displayed in the table, we can draw the following conclusions:

\begin{itemize}
    \item DPN outperforms other baselines on all three public real-world datasets, showcasing its remarkable effectiveness and superiority. In contrast to the leading baseline approach, CAN, DPN exhibited relative performance improvements of 1.53\%, 2.40\%, and 2.95\% on these datasets according to Equation \ref{relaimprov}. Additionally, when compared to the original DIN backbone, DPN displayed even more substantial relative improvements of 6.18\%, 12.72\%, and 6.05\% according to Equation \ref{relaimprov}.
    To delve further into DPN's compatibility with mainstream CTR prediction models, we present compatibility analysis experiments in Section \ref{sec:compatibility}. 
    The superior performance of DPN is attributed to the efficient extraction and sufficient utilization of user behavior patterns, and the follow-up in-depth analysis further demonstrates the significance of user behavior pattern information.
    \item Enhanced modeling of user behavior associations can lead to improved model performance. DIEN and MIMN introduce improvements by modeling the dynamic evolution of user behavior sequences through GRU and Neural Turing Machines. They captured positional relationships between behavior sequences, resulting in enhanced performance. CAN independently model the co-occurrence relationships between the target item and historical interactions on the basis of DIEN, achieving a significant performance boost. These methods have all achieved performance gains through more comprehensive modeling of behavior associations. However, their improvements have remained focused on aggregating behavior representations and modeling co-occurrence relationships between behaviors, while overlooking higher-order behavior pattern information and the pattern-level dependencies. In contrast, DPN efficiently retrieves and fully utilizes target-related user behavior patterns, leading to a remarkable performance improvement
\end{itemize}

\subsection{Ablation Study}

As outlined in Section \ref{sec:Model}, the DPN model comprises three primary components, specifically the TPRM, SPRM, and TPA. To assess the effectiveness of these three elements, we conduct an ablation analysis by eliminating each component from DPN, yielding the subsequent three variants:

\begin{itemize}
    \item \textbf{DPN without TPRM:} Removing TPRM from DPN means no longer retrieving target-related user behavior patterns, i.e., every three sequential user interaction records are considered as behavior patterns to be utilized.
    \item \textbf{DPN without SPRM:} The SPRM in DPN is removed, i.e., the retrieved user behavior patterns are not denoised and are directly utilized.
    \item \textbf{DPN without TPA:} 
    Remove the TPA in DPN, i.e., instead of dependency modeling between the target behavior pattern and the refined historical behavior patterns via attention mechanism, the historical patterns and the target pattern are directly sum-pooled and taken as the feature.
    
\end{itemize}

\begin{table}[!htp]
\caption{Ablation analysis results on three public real-world datasets for three key components in DPN.}
\begin{tabular}{L{2.7cm}C{1.2cm}C{1.2cm}C{1.2cm}}
\toprule
\multirow{2}{*}{Method} & \multicolumn{3}{c}{AUC}  \\\cline{2-4}
                        & Tmall &  Taobao & Alipay  \\\hline
Base Model       & 0.9307& 0.8931& 0.9185\\\hline
DPN w/o TPRM& 0.9507& 0.9308& 0.9371\\
DPN w/o SPRM& 0.9409
& 0.9320
& 0.9364\\
DPN w/o TPA & 0.9456& 0.9312& 0.9347\\\hline
DPN      & 0.9573& 0.9431& 0.9438\\
\bottomrule
\end{tabular}
\label{tab:ablation}
\end{table}

The comparison results on three public real-world datasets are shown in Table \ref{tab:ablation}.
We can draw the following observations from the experiment results:

\begin{itemize}
    \item Evidently, the removal of any individual component within our proposed DPN unequivocally leads to a reduction in performance, thereby substantiating the indispensability and excellence of each constituent element.
    The retrieval, refinement, and dependency modeling of user behavior patterns are all crucial processes for effectively utilizing behavior patterns.
    \item The roles of different modules vary across distinct datasets. The Taobao dataset comprises over a hundred million interaction records, harboring a rich diversity of user interests. Consequently, it is crucial to effectively retrieve target-relevant user behavior patterns within the Taobao dataset, as behavior patterns reflective of unrelated user interests introduce noise into the click-through rate (CTR) prediction task. Conversely, for the Tmall dataset, the refinement of behavior patterns holds paramount importance, possibly due to the greater variety in which these behavior patterns manifest within the interaction records.
\end{itemize}

\subsection{Compatibility Analysis}
\label{sec:compatibility}

DPN is not just a particular model; it's a general framework that can seamlessly work with several well-known CTR prediction models. In order to gauge the compatibility of the DPN framework, we utilize prominent CTR models as the base model within DPN.
We then assess the performance of these resultant models in comparison to the original backbone model. The selected mainstream CTR models for this analysis encompass DNN, DIN, and DIEN, all of which serve as widely employed backbones in various related studies.
The findings from our compatibility analysis experiments are presented in Table \ref{tab:compatibility}.

\begin{table}[!htp]
\caption{Compatibility analysis results on three public datasets with various mainstream CTR prediction methods as the Base Model in DPN.}
\begin{tabular}{L{2.5cm}C{1.4cm}C{1.4cm}C{1.4cm}}
\toprule
\multirow{2}{*}{Method} & \multicolumn{3}{c}{AUC}  \\\cline{2-4}
                        & Tmall &  Taobao & Alipay  \\\hline
DNN       &0.9268   & 0.8731   &   0.9037\\
DPN(DNN)       & 0.9542& 0.9309& 0.9326\\\hline
DIN       & 0.9307   & 0.8931   &   0.9185\\
DPN(DIN)       &0.9573& 0.9431& 0.9438\\\hline
DIEN       & 0.9459   & 0.9058   &   0.9251\\
DPN(DIEN)       & 0.9601& 0.9430& 0.9387\\\hline
\bottomrule
\end{tabular}
\label{tab:compatibility}
\end{table}

According to the experimental results presented in Table \ref{tab:compatibility}, DPN consistently demonstrates significant performance improvements across all backbone models.
The enhancement observed in performance can be attributed to the efficient utilization of rich user behavior pattern information by DPN. The Base Model aggregates user interest representations from historical user interaction records, undertaking the modeling of co-occurrence relationships between the target item and historical interaction records. Building upon this foundation, DPN models user behavior patterns and their inter-relationships, thereby achieving significant performance gains. The aforementioned results underscore the remarkable superiority and broad applicability of DPN, while also providing evidence for the crucial research significance of user behavior pattern mining in CTR prediction tasks.

\subsection{Hyperparameter Analysis}

The key hyperparameters in DPN include: 1) the number of target-related user behavior patterns retrieved in TPRM, i.e., $K$; 2) the length of refined behavior patterns in SPRM, i.e., $s$.
To investigate the impact of these hyperparameters on DPN, we perform a comparison analysis with various settings of the key hyperparameter on public recommendation datasets.
When exploring the interested hyperparameters $K$ and $s$, we keep all other hyperparameters constant.

\begin{figure}[!h]
\centering
\begin{subfigure}{0.327\linewidth}
    \centering
    \includegraphics[width=1.02\linewidth]{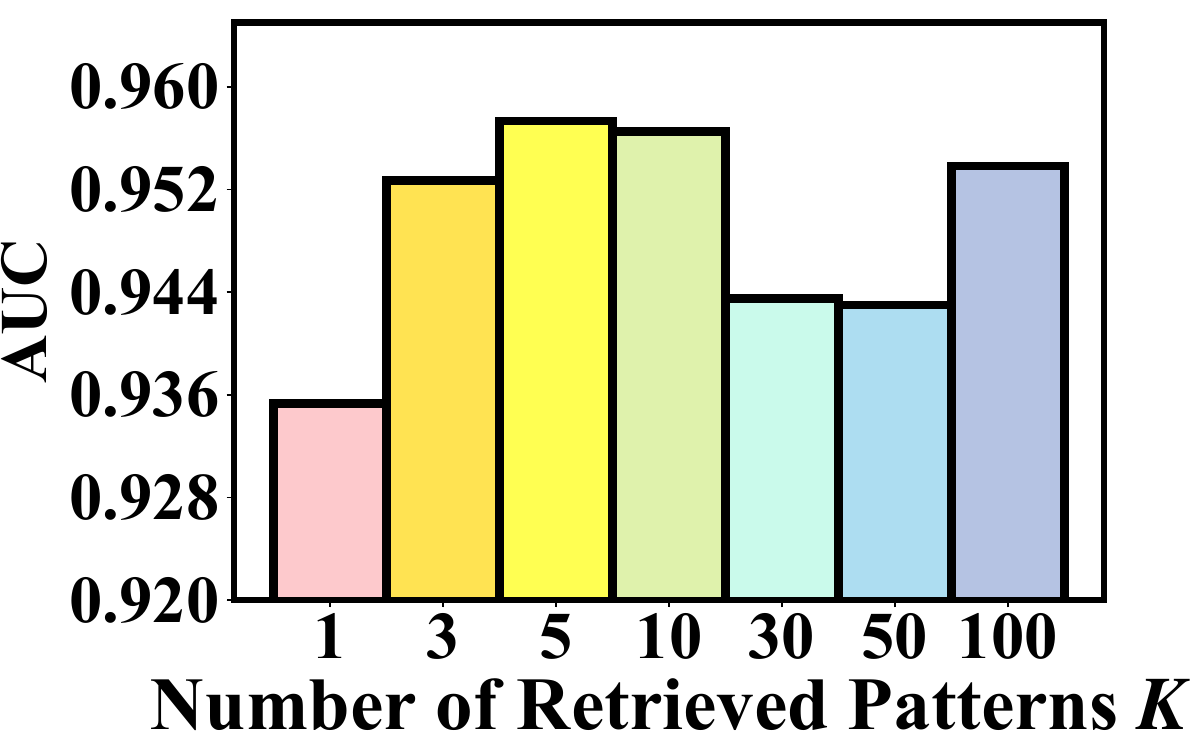} 
    \caption{Tmall}
\end{subfigure}
\begin{subfigure}{0.327\linewidth}
    \centering
    \includegraphics[width=1.02\linewidth]{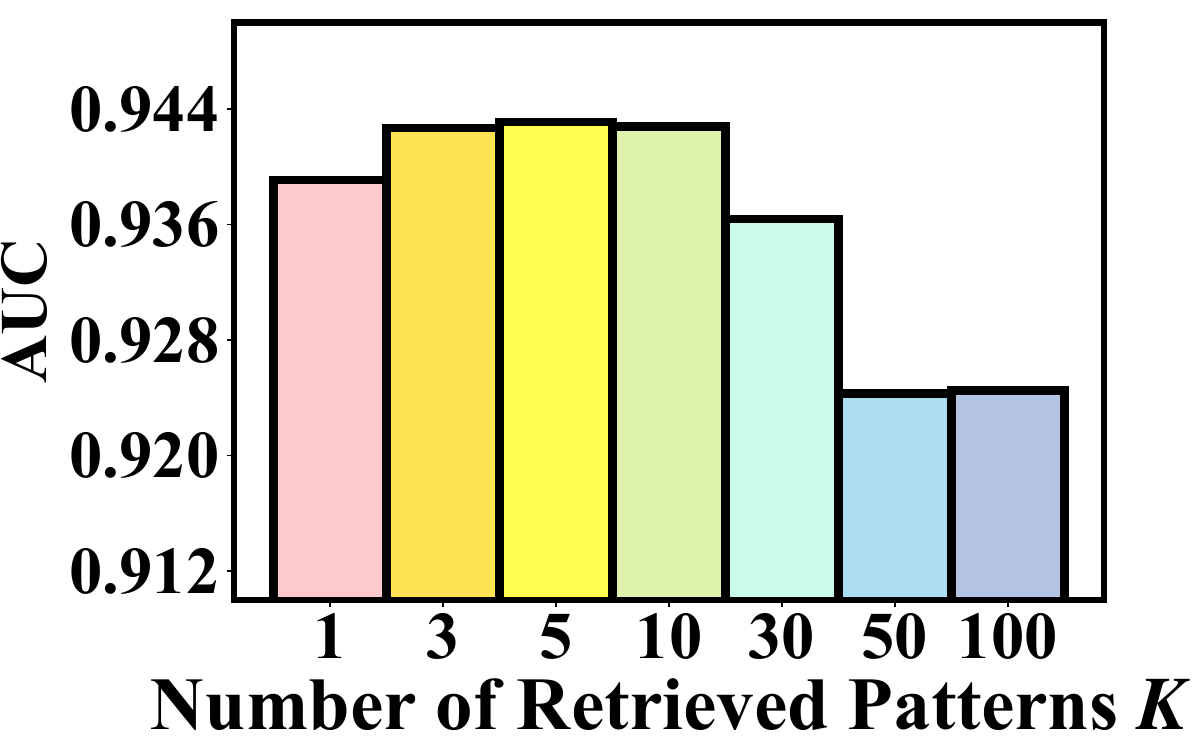} 
    \caption{Taobao}
\end{subfigure}
\begin{subfigure}{0.327\linewidth}
    \centering
    \includegraphics[width=1.02\linewidth]{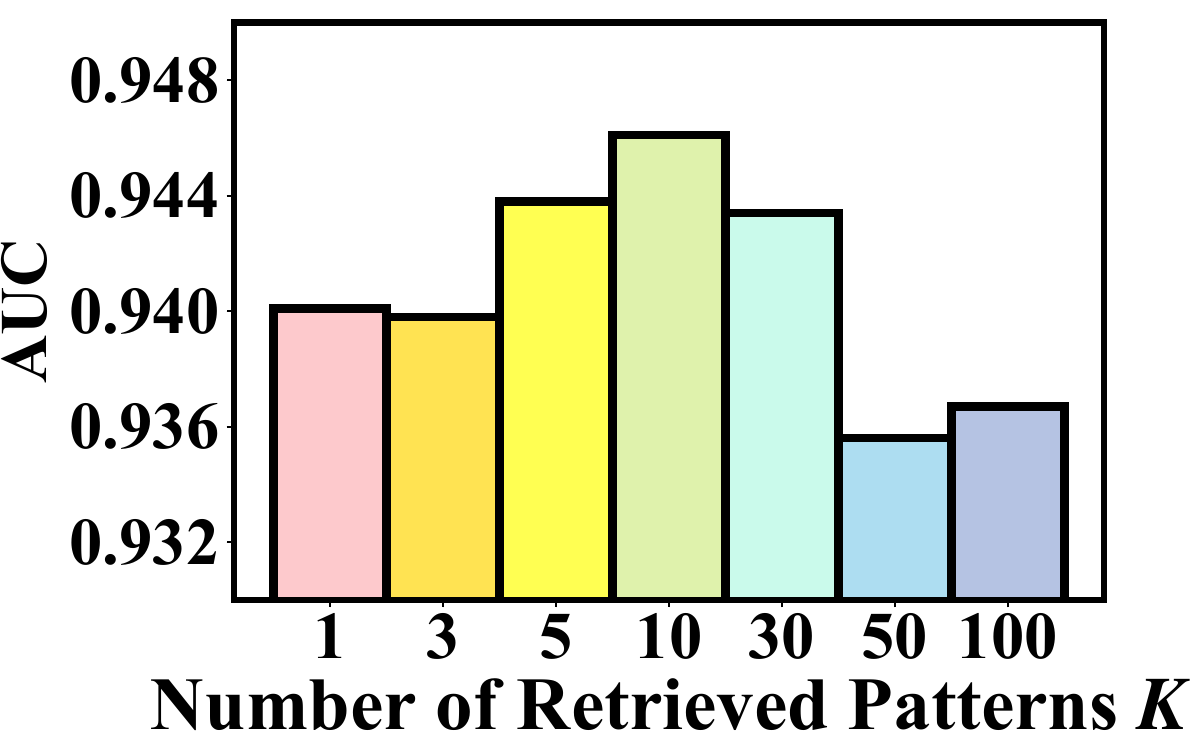} 
    \caption{Alipay}
\end{subfigure}
    \caption{Performances with different numbers of retrieved patterns $K$.}
\label{fig:hyper1}
\end{figure}
\begin{figure}[!h]
\centering
\begin{subfigure}{0.327\linewidth}
    \centering
    \includegraphics[width=1.02\linewidth]{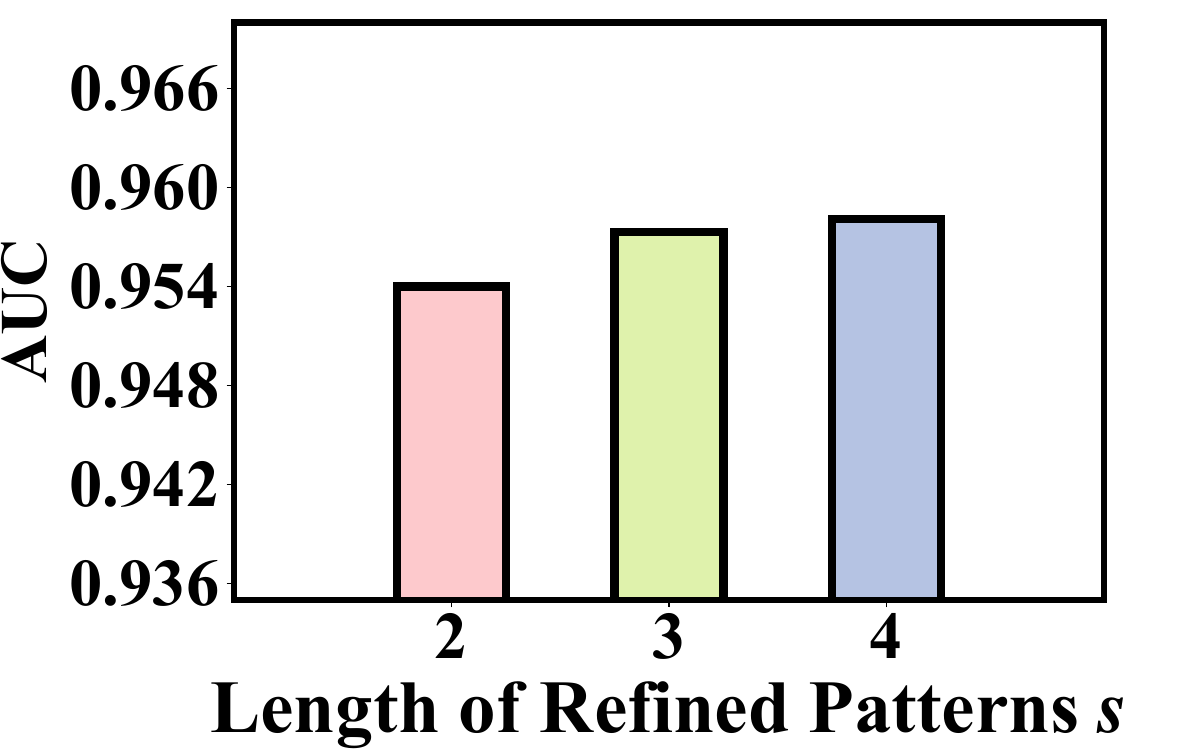} 
    \caption{Tmall}
\end{subfigure}
\begin{subfigure}{0.327\linewidth}
    \centering
    \includegraphics[width=1.02\linewidth]{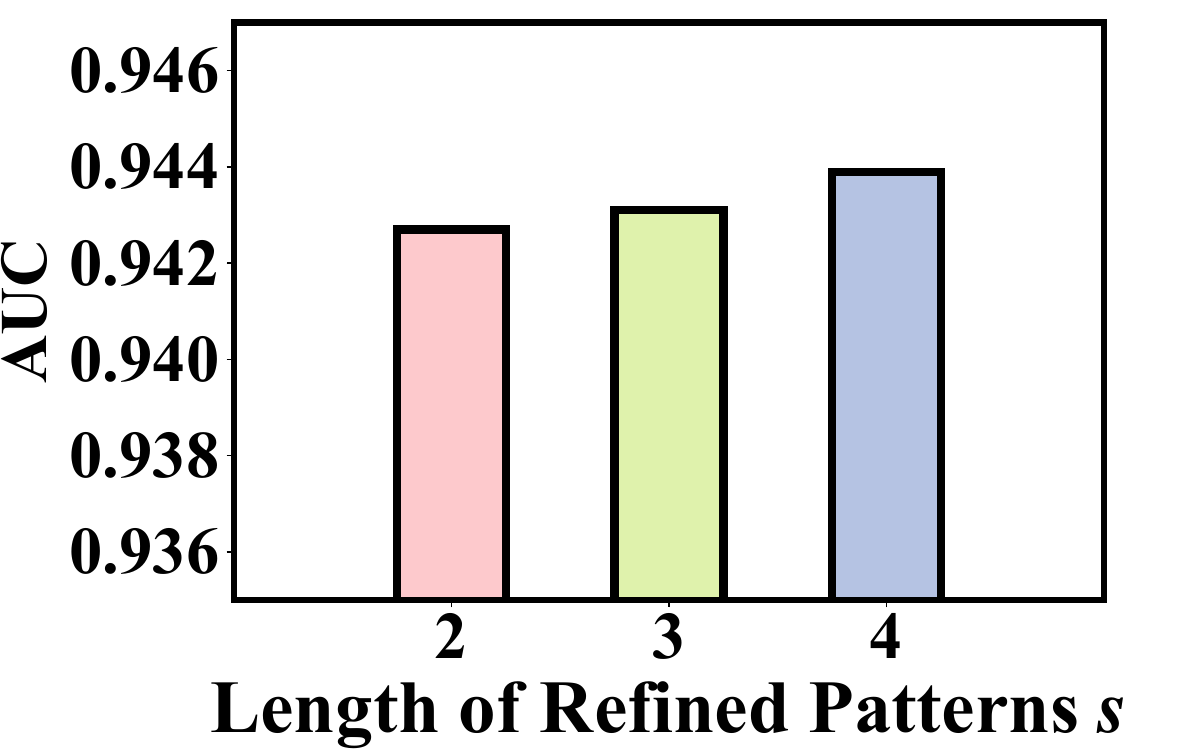} 
    \caption{Taobao}
\end{subfigure}
\begin{subfigure}{0.327\linewidth}
    \centering
    \includegraphics[width=1.02\linewidth]{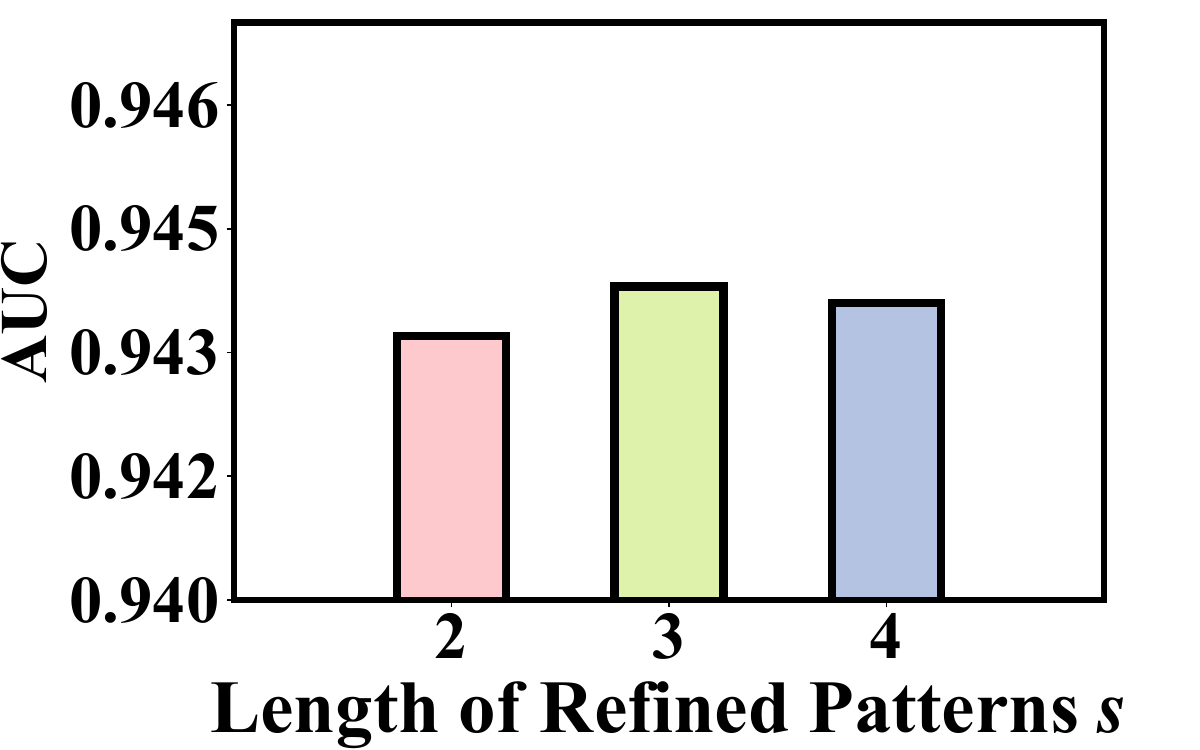} 
    \caption{Alipay}
\end{subfigure}
    \caption{Performances with different lengths of refined patterns $s$.}
\label{fig:hyper2}
\end{figure}

The comparison results with different hyperparameter settings are shown in Figure \ref{fig:hyper1} and \ref{fig:hyper2}.
We can draw the following conclusions:

\begin{itemize}
    \item \textbf{Number of retrieved patterns $K$.} In the case of DPN, optimal performance is achieved when the value of $K$ is moderate across all datasets. This is attributed to the fact that when $K$ is too small, the retrieved user behavior patterns are too limited to adequately reflect the habitual paradigm of users regarding the target item. Conversely, when $K$ is excessively large, the retrieved user behavior patterns become contaminated with numerous target-irrelevant noise patterns, consequently leading to a degradation in the predictive performance of the model.
    \item \textbf{Length of refined patterns $s$.} Generally, higher-order behavior patterns lead to greater performance improvements. This phenomenon can be attributed to two key factors. On one hand, higher-order behavior patterns to some extent reflect information from lower-order patterns. On the other hand, higher-order patterns possess unique pattern characteristics, enabling them to achieve superior performance compared to lower-order patterns. However, higher-order patterns do not always yield better results. For instance, when comparing a 4-order DPN to a 3-order DPN on the Alipay dataset, there was a slight decrease in performance. This could be attributed to the dataset's lack of high-order behavior patterns. Therefore, an increase in the length of behavior patterns did not introduce valuable pattern characteristics and potentially introduced unwanted noise.
\end{itemize}

\subsection{Case Study}

\begin{figure}[!ht]
    \centering
    \includegraphics[width=0.98\linewidth]{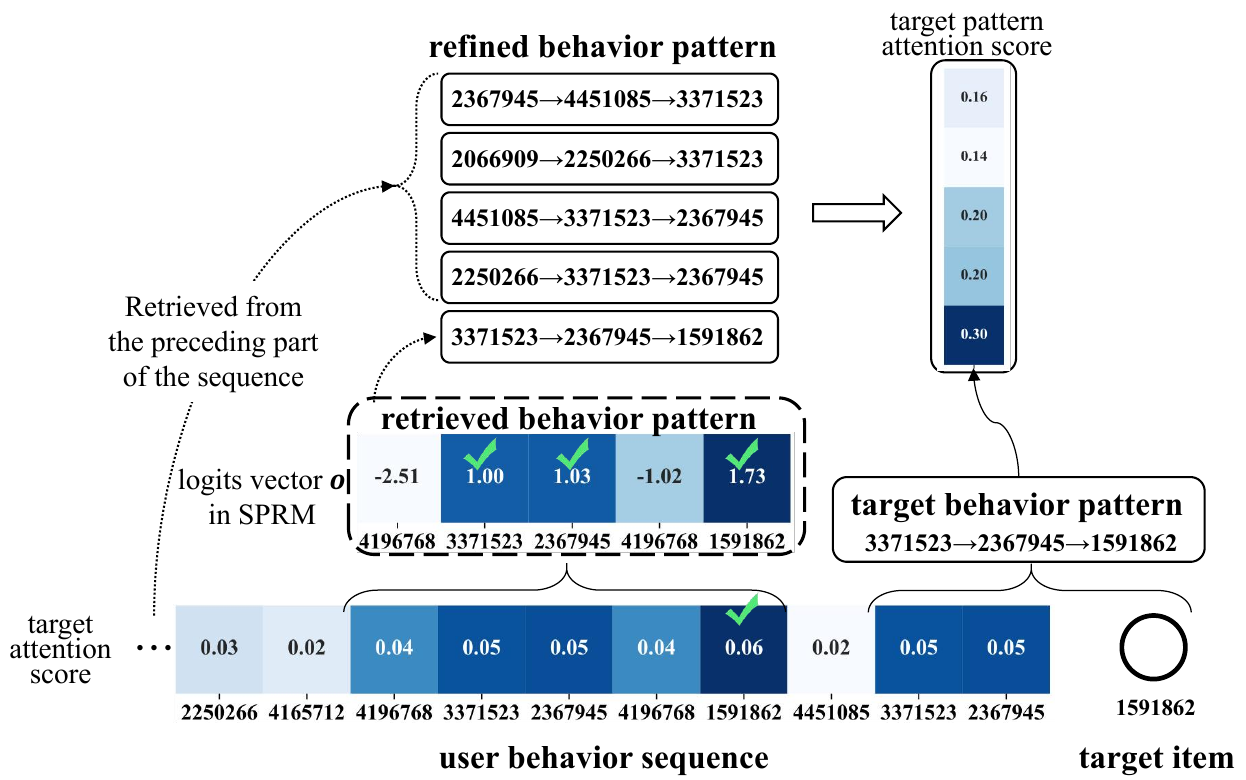}
    \caption{Illustration of Case Study on DPN. This figure visualizes the process of DPN handling a real-world user behavior sequence. For simplification, only the last 10 behavior records are shown and the preceding part is omitted. The numbers under the heatmap are the item IDs.}
\label{fig:case}
\end{figure}

To validate that DPN operates as designed, we employ a case study approach to elucidate the internal structure of DPN.
Figure \ref{fig:case} illustrates the process of handling a real-world instance in DPN, where TPRM retrieves target-relevant behavior patterns, SPRM further refines the retrieved patterns and TPA aggregates the refined behavior pattern representations while capturing the dependency among patterns.

In this case, the target behavior pattern is $\textit{3371523}\rightarrow \textit{2367945}\rightarrow \textit{1591862}$, which is exactly the most high-frequency behavior pattern of Taobao shown in Figure \ref{fig:intro}. 
Firstly, TPRM retrieves the Top-5 target-related behavior patterns, and the last retrieved one is $\textit{(4196768, 3371523, 2367945, 4196768, 1591862)}$.
Subsequently, the SPRM further refines this pattern, presenting the refined behavior pattern $\textit{3371523}\rightarrow \textit{2367945}\rightarrow \textit{1591862}$, and it is identical to the target pattern. 
Finally, the TPA models the dependency between the target pattern and refined patterns via the attention mechanism. 

From the insights in the case study, DPN seems to handle behavior patterns rightly. 
To further analyze the effectiveness of DPN, we design experiments to discuss the ability of DPNs to refine behavior patterns and model dependencies between behavior patterns by means of a more global visualization in the following Subsections \ref{sec:refine} and \ref{sec:mi}.
We did not delve extensively into pattern retrieval due to space constraints, as it primarily aims to avoid the computational complexity introduced by behavior pattern modeling, which is not the core focus of our work. The core of our work lies in learning pattern information and modeling dependencies between patterns.

\subsection{Validation of Pattern Refinement}
\label{sec:refine}

The actions of the user are driven by diverse interest.
So the behavior patterns driven by different interest may be mutually permeated, and not be perfectly continuous segments.
Moreover, misclick is also not a rare event, which introduces noise into behavior patterns.
To capture more meaningful patterns and mitigate the risk of introducing noise, we design SPRM to extract $s$-length behavior patterns from the retrieved behavior patterns.
Meaningful patterns refer to which have stronger intra-dependencies, which typically can be measured by the concept of mutual information (MI).
To validate the effectiveness of refinement, we define Intra-Pattern Mutual Information (IntraPMI) to evaluate the intra-dependencies within the behavior pattern, referencing~\cite{tc} which extends MI to the multi-variables case.
IntraPMI measures all the possible components of correlations among the behaviors within the behavior pattern, so a pattern with higher IntraPMI is a more meaningful pattern.

\begin{figure}[!h]
    \centering
    \includegraphics[width=1.02\linewidth]{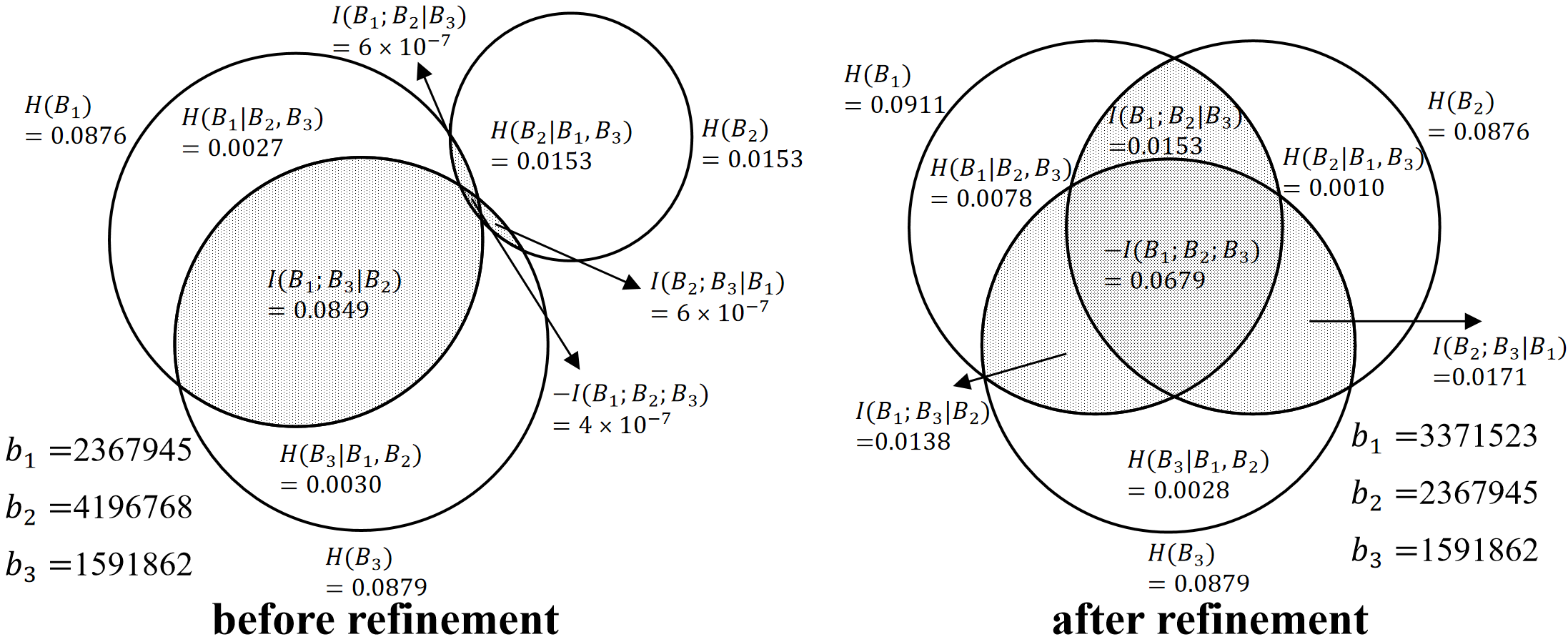} 
    \caption{ Venn diagram of information measurement among items in behavior patterns before/after refinement (using the same case in Figure \ref{fig:case} as an example). The dotted area corresponds to the IntraPMI, reflecting the dependencies among the items. It is obvious that the intra-pattern dependency is more closely knit after refinement. }
\label{fig:venn_refine}
\end{figure}

\begin{definition}
    \textbf{Intra-Pattern Mutual Information.}
    The intra-pattern mutual information of a given behavior pattern $p=\{b_1, b_2, b_3\}$ is defined as:
    \begin{equation}
    \begin{aligned}
        \mathop{IntraPMI}(p) = H(B_1)+H(B_2)+H(B_3)-H(B1;B2;B3), \\ B_i=\{0, 1| \mathbb{I}(b_i\text{ interacted by the user})\}
        \end{aligned}
    \end{equation} 
    where $B_i$ is the binary variable indicating whether the user interacts with $b_i$, $\mathbb{I}(\cdot)$ is the indicator function that outputs 1 when the condition is valid and 0 vice versa.
    $H(B_i)$ means the information entropy of variable $B_i$, and $H(B1;B2;B3)$ is the joint entropy of variables $B_1, B_2 ,B_3$.

\end{definition}

To illustrate what IntraPMI means, a Venn diagram of information measurement among items within behavior patterns before/after refinement is shown in Figure \ref{fig:venn_refine}.
IntraPMI corresponds to the dotted area (the overlap of circles) in Figure \ref{fig:venn_refine}, which measures the amount of all the correlations among the behaviors within the pattern.
According to Figure \ref{fig:venn_refine}, the IntraPMI of the pattern after refinement is significantly larger, i.e., the circles corresponding to items are more closely knit, showing more stronger intra-pattern dependencies.

\begin{figure}[!h]
\centering
\begin{subfigure}{0.4955\linewidth}
    \centering
    \includegraphics[width=1.02\linewidth]{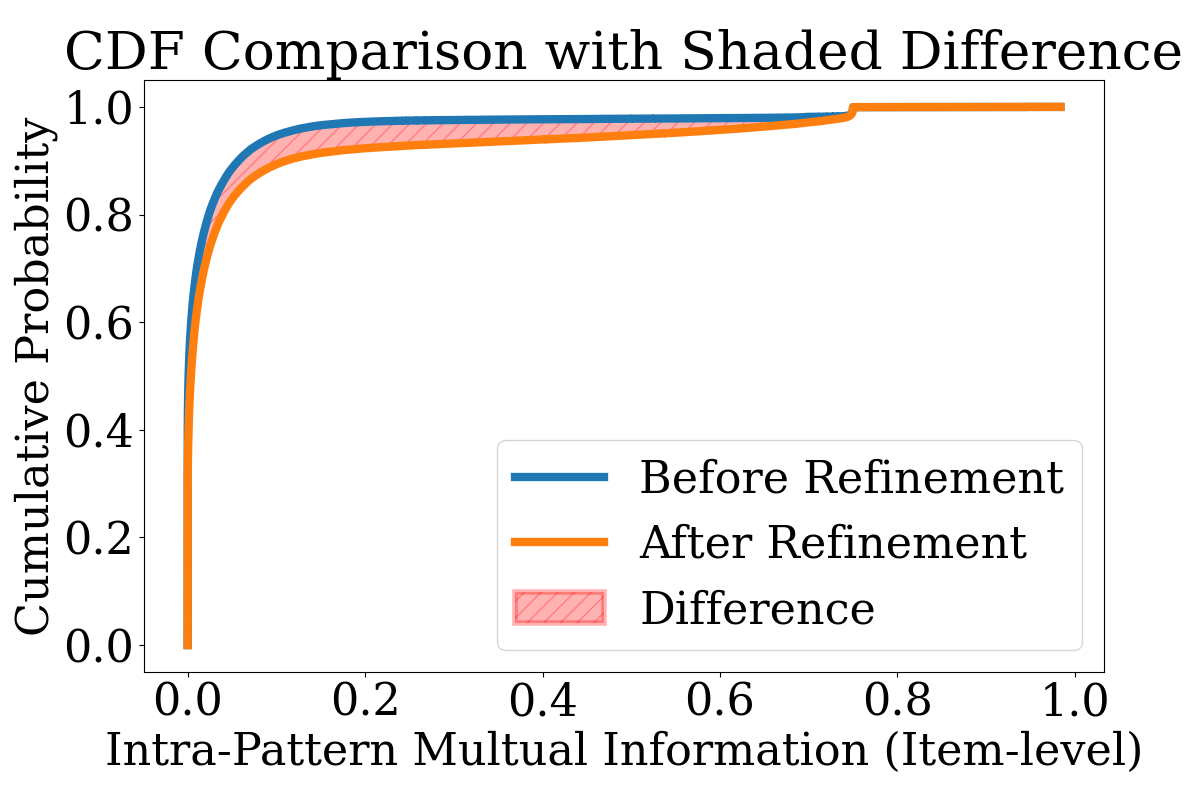} 
\end{subfigure}
\begin{subfigure}{0.4955\linewidth}
    \centering
    \includegraphics[width=1.02\linewidth]{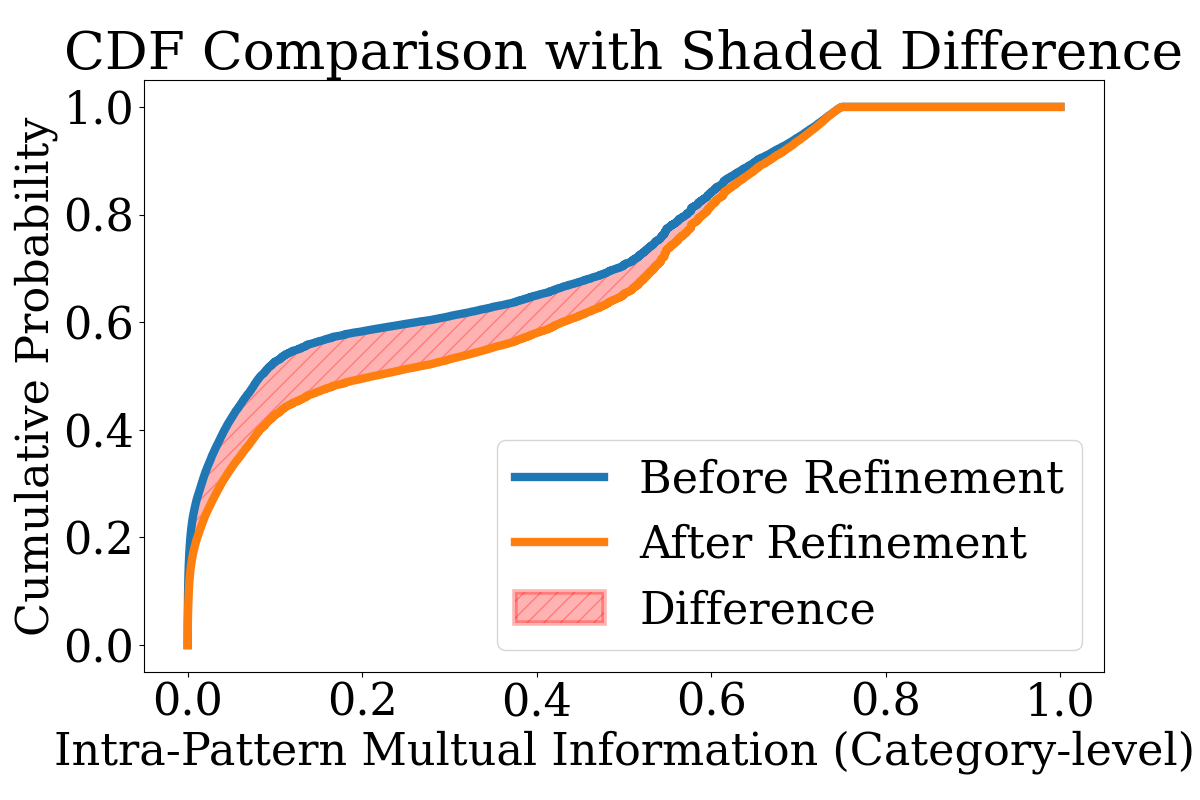} 
\end{subfigure}
    \caption{Cumulative Distribution Function (CDF) of the Intra-Pattern Mutual Information of patterns before/after refinement on both item and category level. The difference in CDFs is indicated by the red diagonal texture.}
\label{fig:cdf}
\end{figure}

To conduct a more general analysis instead of a case study, we visualize the Cumulative Distribution Function (CDF) of the IntraPMI of patterns before/after refinement on the Taobao dataset as shown in Figure \ref{fig:cdf}. 
Notably, we normalize the metric to range [0, 1] following~\cite{msu}.
The CDF curves corresponding to the refined patterns are located below the ones before refinement, indicating that our proposed SPRM can make the mutual information of the patterns larger, i.e. more meaningful.

\subsection{What does DPN capture?}
\label{sec:mi}

To investigate which dependencies among patterns are valuable for the CTR prediction task and what models can capture, we use the concept of conditional mutual information which is widely used in feature selection~\cite{feature_select1, feature_select2} to reflect the correlation between features and labels for visualization.

\subsubsection{Metrics}
For a given historical behavior pattern $p_i$, whether $p_i$ occurs in the user behavior sequence can be described as a binary variable $\mathcal{P}_H^{(p_i)}=\{0, 1| \mathbb{I}(p_i\text{ occurs in the user behavior records})\}$, where $\mathbb{I}(\cdot)$ denotes the indicator function to discriminate the condition is true or false.
Similarly, $\mathcal{P}_T^{(p_t)}=\{0, 1| \mathbb{I}(p_t\text{ is the target pattern})\}$ indicates whether the behavior pattern $p_t$ is the target pattern.

The conditional mutual information $I(\mathcal{P}_H^{(p_i)}; \mathcal{Y}|\mathcal{P}_T^{(p_t)})$ measures the dependency relationship between $\mathcal{P}_H^{(p_i)}$  and the labels $\mathcal{Y}$ given $\mathcal{P}_T^{(p_t)}$, reflecting the value of historical pattern $p_i$ for CTR prediction under a given target pattern $p_t$.
It is the ground truth that demonstrates which dependencies among patterns are valuable for the CTR prediction.

The conditional mutual information $I(\mathcal{P}_H^{(p_i)}; \mathcal{\hat Y}|\mathcal{P}_T^{(p_t)})$ measures the dependency relationship between $\mathcal{P}_H^{(p_i)}$  and the click probability $\mathcal{\hat Y}$ predicted by the model given $\mathcal{P}_T^{(p_t)}$.
$I(\mathcal{P}_H^{(p_i)}; \mathcal{\hat Y}|\mathcal{P}_T^{(p_t)})$ on test dataset reflects the effect of pattern-level dependencies on CTR prediction learned by the model.

\begin{figure}[!h]
\centering
\begin{subfigure}{0.4955\linewidth}
    \centering
    \includegraphics[width=1.02\linewidth]{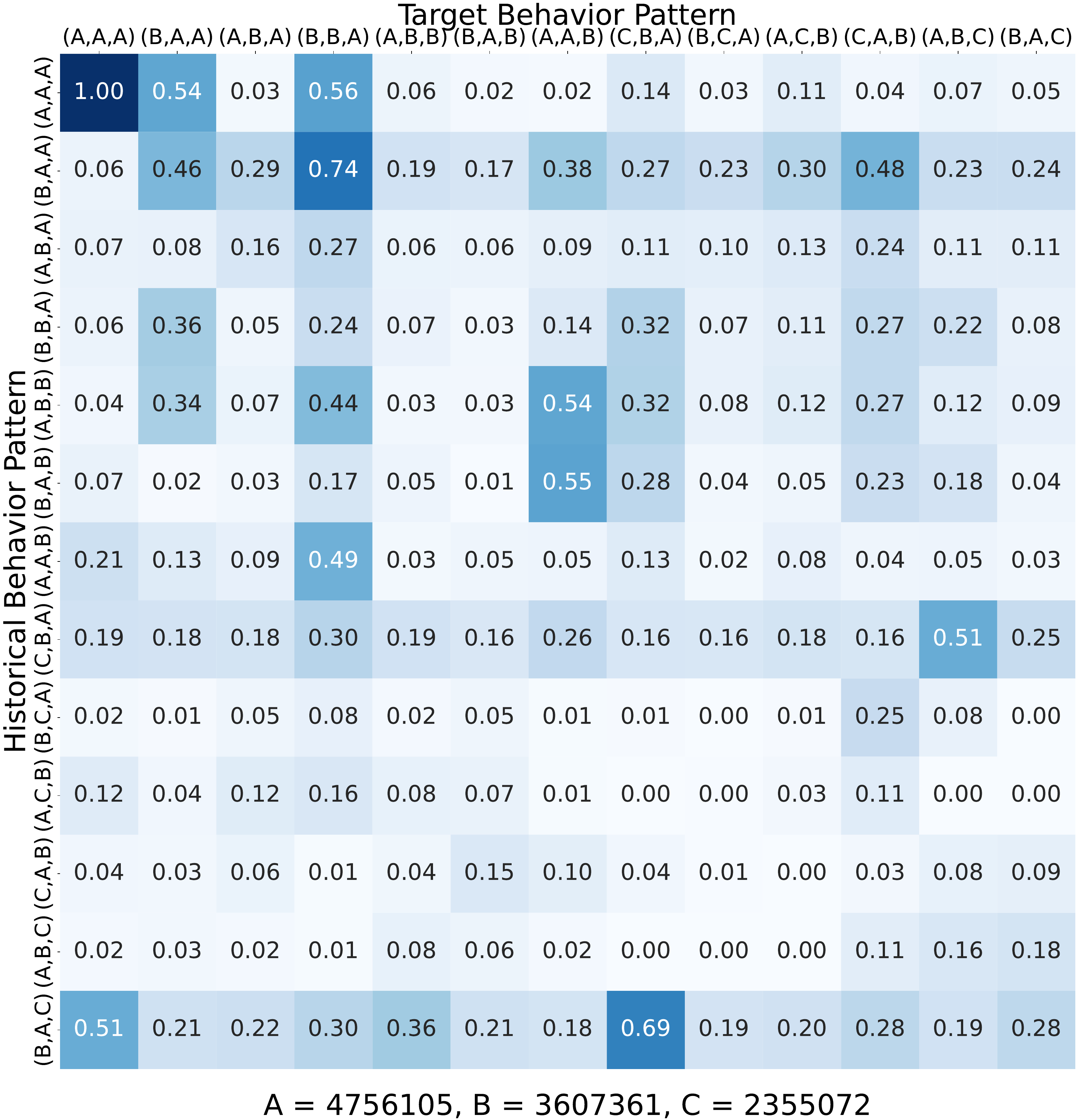}
    \caption{Ground truth}
\end{subfigure}
\begin{subfigure}{0.4955\linewidth}
    \centering
    \includegraphics[width=1.02\linewidth]{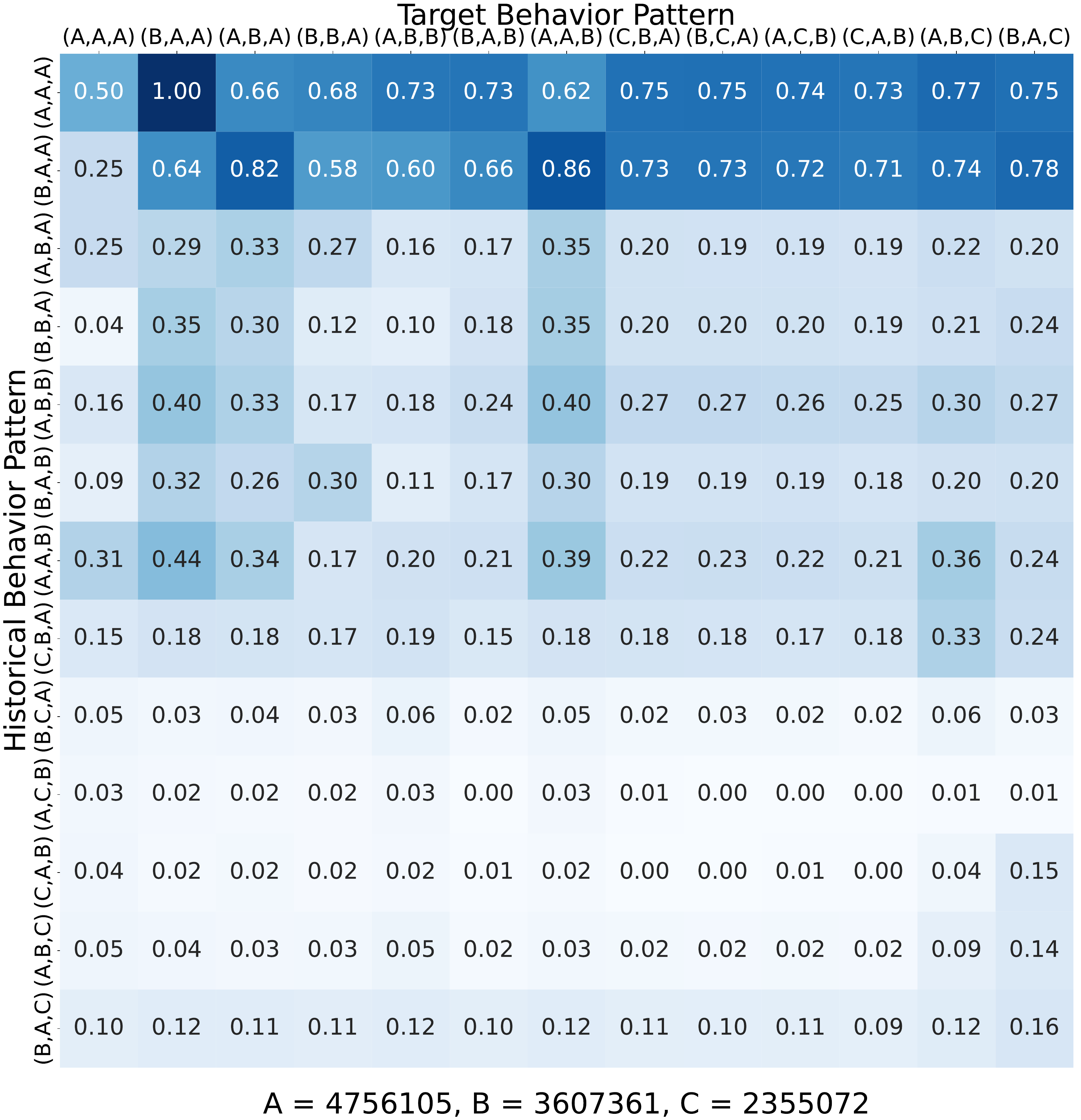} 
    \caption{DIN}
\end{subfigure}
\begin{subfigure}{0.4955\linewidth}
    \centering
    \includegraphics[width=1.02\linewidth]{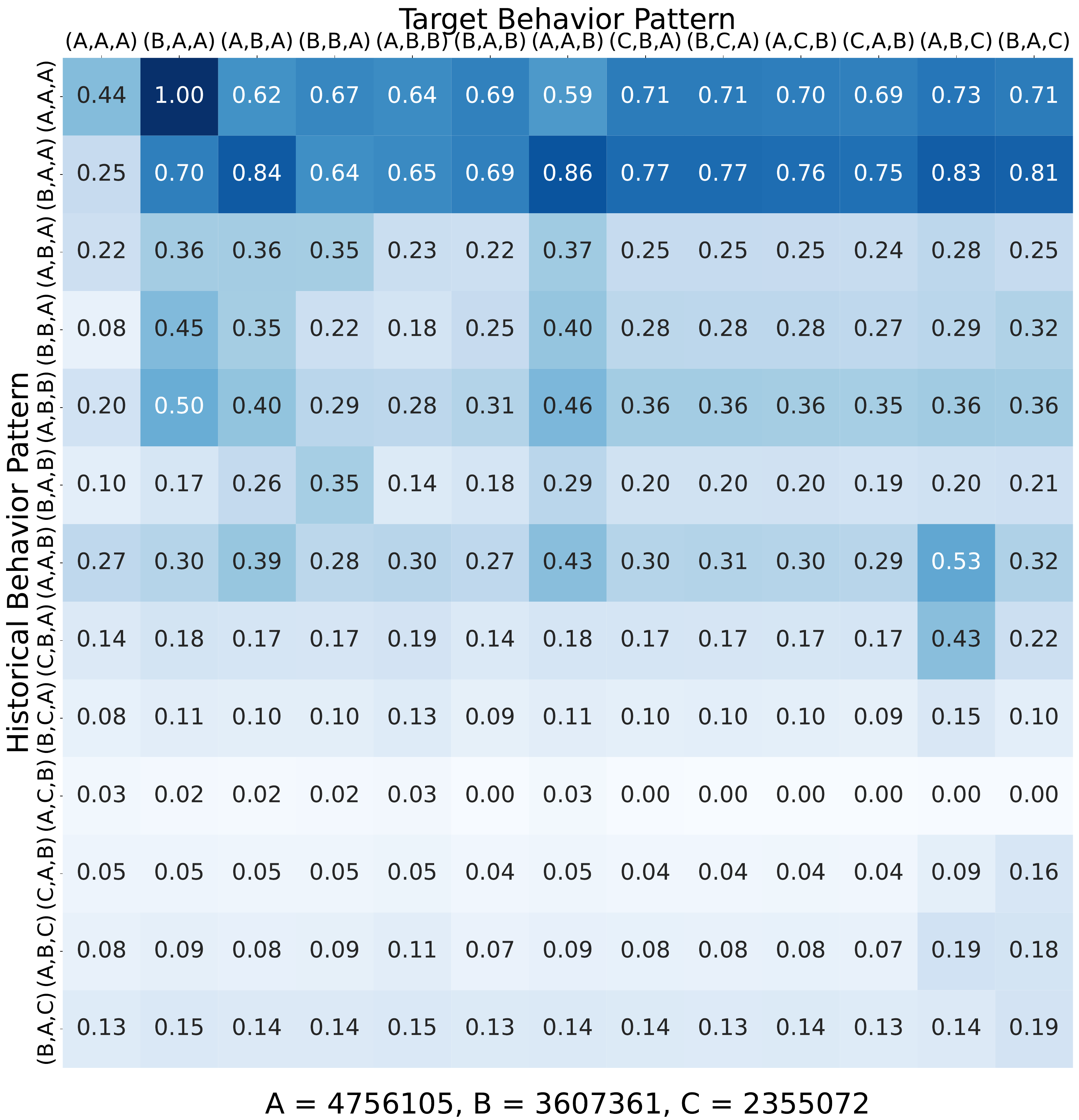}
    \caption{DIEN}
\end{subfigure}
\begin{subfigure}{0.4955\linewidth}
    \centering
    \includegraphics[width=1.02\linewidth]{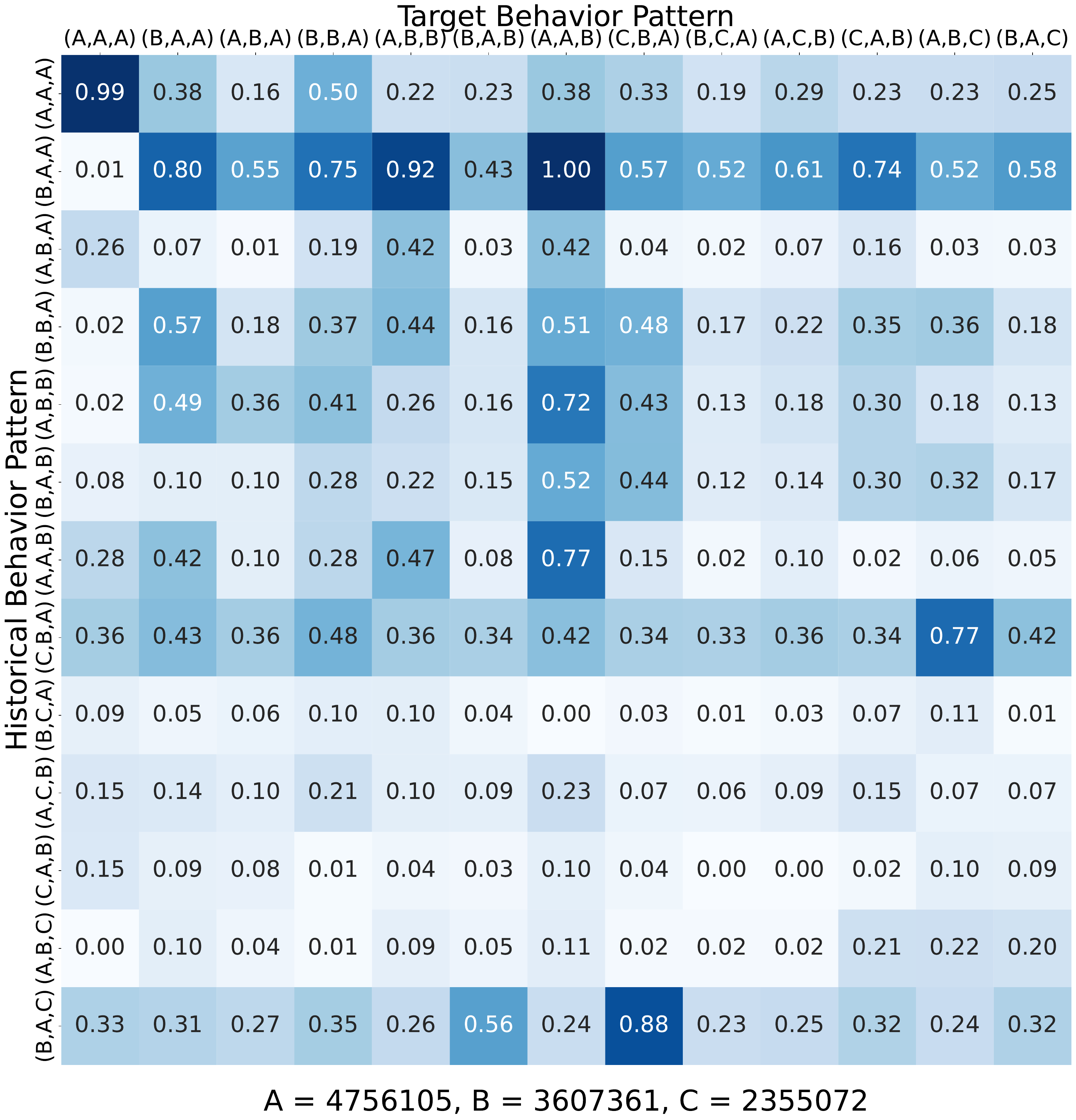} 
    \caption{DPN}
\end{subfigure}
    \caption{Visualization of the conditional mutual information.}
\label{fig:mi}
\end{figure}

\subsubsection{Analysis}
We select the Top-3 frequent categories 4756105, 3607361, and 2355072 (denoted as A, B, C for simplification) in the test dataset of Taobao dataset for analysis.
There are many possible patterns that can be formed by A, B, and C.
We visualize $I(\mathcal{P}_H^{(p_i)}; \mathcal{Y}|\mathcal{P}_T^{(p_t)})$ for some of them as shown in the Figure \ref{fig:mi} (a).
It is obvious that some dependency between the historical patterns and the target pattern is meaningful (with a higher value in the matrix) to CTR prediction tasks, such as (A, A, A) \& (A, A, A), (B, A, A) \& (B, B, A) and (B, A, C) \& (C, B, A).

Furthermore, we visualize $I(\mathcal{P}_H^{(p_i)}; \mathcal{\hat Y}|\mathcal{P}_T^{(p_t)})$ for the DPN and the most respective behavior-based CTR prediction models, i.e., DIN and DIEN on the test dataset of Taobao dataset as shown in Figure \ref{fig:mi} (b-d).
According to the visualization results, we can draw the following conclusions:

\begin{itemize}
    \item DPN can capture the valuable pattern-level dependency well. (a) and (d) in Figure \ref{fig:mi} exhibit a noticeable correlation. On the other hand, DIN and DIEN are completely unable to do so, as (a) is significantly different from (b)/(c).
    \item DIN and DIEN are not aware of the target pattern. For example, regarding target patterns with the target category A in Figure \ref{fig:mi} (ending with A, such as (B, A, A), (A, B, A), (B, B, A), (C, B, A), (B, C, A)), their corresponding columns in the conditional mutual information matrix exhibit strong cosine similarity (>0.9) pairwise.
    \item DIN and DIEN are not sensitive to the historical patterns. For the patterns consisting of the same items, such as (B, A, A), (A, B, A), (A, A, B), their corresponding rows in the matrix are obviously similar. DIN can not handle the sequential dependency, so it can not capture the pattern information. DIEN handles the user behavior sequence via a unified sequential model, modeling the global evolution of user interest while neglecting the local behavior patterns.

\end{itemize}

In summary, DPN not only learns the pattern information but also captures the valuable dependencies among behavior patterns, thus achieving significant performance gain for the CTR prediction.

\section{Conclusions}

In this article, we highlight the significant importance of diverse user behavior patterns hidden within massive historical interaction records and the neglect of this idea by existing CTR prediction methods. 
To fully leverage user behavior pattern information, we propose the Deep Pattern Network (DPN) for click-through rate prediction.
DPN not only learns the information of behavior patterns but also captures the valuable dependencies among behavior patterns.
Comprehensive experiments conducted on three different datasets thoroughly demonstrate the outstanding superiority and broad compatibility of our proposed DPN. Further analysis elucidates how the rich behavior patterns enhance the performance of DPN.

\bibliographystyle{ACM-Reference-Format}
\balance
\bibliography{ref}

\end{document}